\documentclass[traditabstract,useAMS,usenatbib]{aa} 
\usepackage{natbib,graphicx}
\usepackage{times}
\usepackage{txfonts}
\usepackage{color}                
\def\kms{km ${\rm s}^{-1}$}

\def\ch2{$\chi^2$}
\def\dg{$^{\circ}$}

\def\kms {\hbox{${\rm km\ s}^{-1}$}}

\def\ccm {$\hbox{{\rm cm}}^{-3}$}    

\def\MOLH {\hbox{${\rm H}_2$}}  

\def \AL {$\alpha $}     
\def \HI {H{\sc \,i}}
\def \carbi {\hbox{[C{\sc \,i}]}}
\def \carbii {\hbox{[C{\sc \,ii}]}}

\def\lapp{\ifmmode\stackrel{<}{_{\sim}}\else$\stackrel{<}{_{\sim}}$\fi}
\def\gapp{\ifmmode\stackrel{>}{_{\sim}}\else$\stackrel{>}{_{\sim}}$\fi}
\begin{document}
%
\title{Measuring  space-time variation of the fundamental constants with redshifted submillimetre transitions of neutral carbon}
 \subtitle{}
   \author{S. J. Curran\inst{1}, A. Tanna\inst{1},  F. E. Koch\inst{1}, J. C. Berengut\inst{1}, J. K. Webb\inst{1}, A. A. Stark\inst{2}, V. V. Flambaum\inst{1}
     }

   \institute{School of Physics, University of New
  South Wales, Sydney NSW 2052, Australia\\
  \email{sjc@phys.unsw.edu.au}
  \and
  Smithsonian Astrophysical Observatory,   60 Garden Street,  Cambridge  MA 02138, U.S.A.
         }
            
   \date{}

 
   \abstract{We compare the redshifts of neutral carbon and carbon monoxide in the redshifted sources in
     which the $^3{\rm P}_{1}\rightarrow\,^3{\rm P}_{0}$ fine structure transition of neutral carbon, \carbi,  has been detected,
     in order to measure space-time variation of the fundamental constants.  Comparison with the CO rotational lines
     measures $F\equiv\alpha^2/\mu$, where \AL\ is the fine structure constant and $\mu$ is the electron-proton mass
     ratio, which is the same combination of constants obtained from the comparison
$^2{\rm P}_{3/2}\rightarrow\,^2{\rm P}_{1/2}$ fine structure line of singly ionised
carbon, \carbii. However, neutral carbon has the distinct advantage that it may be spatially coincident with the carbon monoxide,
whereas \carbii\ could be located in the diffuse medium between molecular clouds, and so any comparison with CO 
could be dominated by intrinsic velocity differences. Using \carbi, we obtain a mean variation of
$<\Delta F/F> = (-3.6\pm 8.5)\times10^{-5}$, over $z = 2.3 - 4.1$, for the eight\carbi\ systems, which degrades
to $(-1.5\pm 11)\times10^{-5}$, over $z = 2.3 - 6.4$ when the two \carbii\ systems are included. That is, zero
variation over look-back times of 10.8--12.8 Gyr. However, the latest optical results indicate a spatial variation in \AL,
where $\Delta \alpha/\alpha$ describes a dipole and we see the same direction in $\Delta F/F$. This trend is, however, due to a single source for which the \carbi\ spectrum is
of poor quality. This also applies to one of the two \carbii\ spectra previously used to find a zero variation in
$\alpha^2/\mu$. Quantifying this, we find an anti-correlation between $|\Delta F/F|$ and the quality of the carbon detection,
as measured by the spectral resolution, indicating that the typical values of $\gapp50$ \kms, used to obtain a detection, are too coarse
to reliably  measure changes in the constants. From the fluxes of the known $z\gapp1$ CO systems, we predict that 
{\em current instruments are incapable of the sensitivities required to measure changes in the constants through the
comparison of CO and carbon lines}.
We therefore discuss in detail the use of ALMA for such an 
undertaking and find that, based upon the current CO detections {\em only},  the {\em Full Array} configuration is expected to detect $\sim100$ galaxies in
\carbi\ at better than 10 \kms\ spectral resolution, while potentially resolving the individual molecular cloud
complexes at redshifts of $z\gapp3$. This could provide $\gapp1000$  individual systems with which to obtain
accurate measurements of space-time variation of the constants at look-back times in excess of 11 Gyr. 

   }

\authorrunning{S. J. Curran et al.}
\titlerunning{Carbon measurements of space-time variation of constants}
   \maketitle
%
\section{Introduction}

Atomic carbon is an important tracer of the dense neutral gas within the Galactic and extragalactic interstellar
media. With rest frequencies of $\nu_{\rm rest} = 492.160651 (55)$ GHz \citep{ys91a}
 and $\nu_{\rm rest} = 809.34197 (5)$ GHz \citep{kls+98} for the $^3{\rm
  P}_{1}\rightarrow\,^3{\rm P}_{0}$ and $^3{\rm
  P}_{2}\rightarrow\,^3{\rm P}_{1}$ fine structure transitions, respectively, neutral carbon is difficult to observe at low redshift,
although submillimetre observations of \carbi\ in the Galaxy \citep{osh+01}, M82 \citep{wec+94,sgh+97} and
other near-by galaxies \citep{gp00,ib02}, show that the \carbi\ emission is closely associated with that of carbon
monoxide. This applies over a wide range of different environments and is most likely due
to their similar critical densities ($n\sim10^3$ \ccm). Furthermore, it has been suggested that \carbi\ can provide a measure of
the \MOLH\ gas mass independent of CO \citep{pg04,ptv04}, the standard molecular hydrogen tracer.

Although the neutral carbon is generally believed to be located on the surfaces of molecular
clouds,  it can be found deep inside the cloud \citep{iot+02}, giving the spatial coincidences
mentioned above and
making the comparison of \carbi\ and CO redshifts
potentially very useful in measuring the values of the fundamental constants at large look-back times.
Comparison of the fine structure lines with the CO rotational lines gives $\Delta F/F$,
where $F= \alpha^2/\mu$, with \AL\ being the fine structure constant\footnote{$\alpha\equiv
    e^2/\hbar c$, where $e$ is the charge on the electron, $\hbar$ Planck's constant/$2\pi$ and $c$ the speed of
    light.} and  $\mu = m_{\rm e}/m_{\rm p}$ the electron-proton mass ratio.  Relative shifts in the millimetre-wave band are more than an
order of magnitude larger than in the optical (e.g. \citealt{mwf+00}), from which the intra-comparison of redshifted
ultra-violet electronic transitions, in the absorption spectra of metal-ions in damped Lyman-alpha
  systems, indicate that \AL\ has changed over the history of the Universe \citep{mwf03}.

  As well as a temporal variation having implications for current grand unified theories, a spatial variation of the
  constants may address the prominent ``fine tuning'' issue, where if the fundamental constants (which are dependent
  upon 27 independent parameters) were even slightly different than the observed values, life could not appear.
Such a spatial variation may recently have been detected by \citet{wkm+10}, where \AL\ has a different
values in different regions of the sky, described by a dipole (see also \citealt{bfk+10}).

Thus, redshifted \carbi\ lines are of great interest in measuring the space-time variation of the fundamental constants, especially since
the comparison CO line has already been detected in over a hundred cases at $z\gapp1$ 
(see Sect. \ref{pci}), a number which will greatly increase with the advent of the Atacama Large Millimetre Array (ALMA).

\citet{lrk+08} have already compared the $^2{\rm P}_{3/2}\rightarrow\,^2{\rm P}_{1/2}$ fine structure line of singly ionised
carbon\footnote{$\nu_{\rm rest} = 1900.5369 (1.3)$ GHz \citep{cbs86}.},  \carbii, with the rotational transitions of the
CO lines, which also gives $F= \alpha^2/\mu$, at $z = 4.69$ and $z=6.42$.\footnote{From CO and \carbii\ in the quasars
BR\,1202--0725 \citep{opg+96,iye+06} and J1148+5251 \citep{bcn+03,mcc+05}.}  They find no temporal variation, although the
uncertainties are large due to the weak \carbii\ spectra, which are very difficult to detect at high redshift (see
\citealt{cur09}). Furthermore, unlike the neutral carbon, ionised carbon traces photo-dissociation regions in the diffuse
interstellar medium between molecular clouds and is therefore unlikely to be spatially coincident with the CO.
 In any case, a non-detected temporal variation does not necessarily imply a zero
spatial variation, which may possibly account for discrepancies which arise in changes in \AL\ when attributed
purely to a temporal variation (e.g. \citealt{lcm+06}).

In this paper, we fit the observed \carbi\ and CO  profiles of the known redshifted systems in
order to derive the difference in redshifts between the species in each of the sources, thus yielding
the values of $\alpha^2/\mu$ at the redshifts and locations of these quasars. 


\section{Analysis}
\subsection{Profile fitting}

\begin{figure*}
\vspace{15.0cm}  
\includegraphics{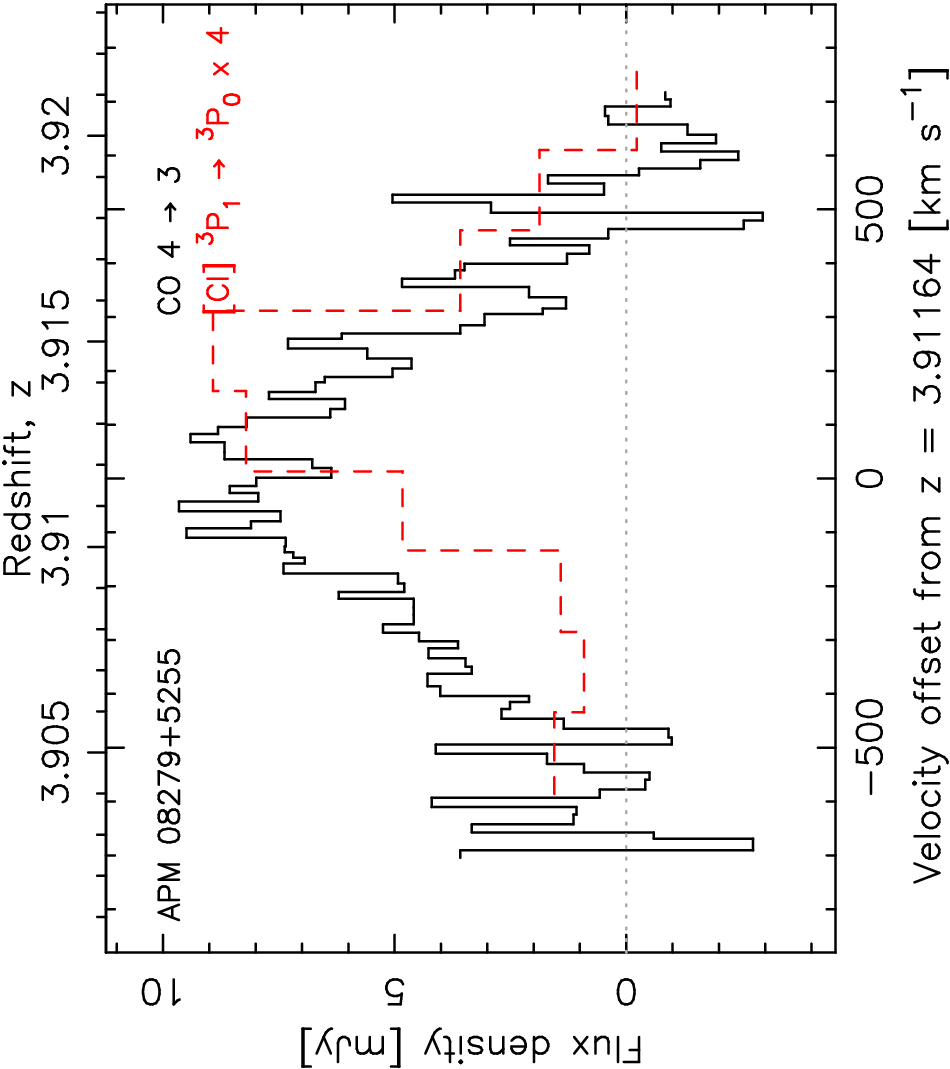}
\includegraphics{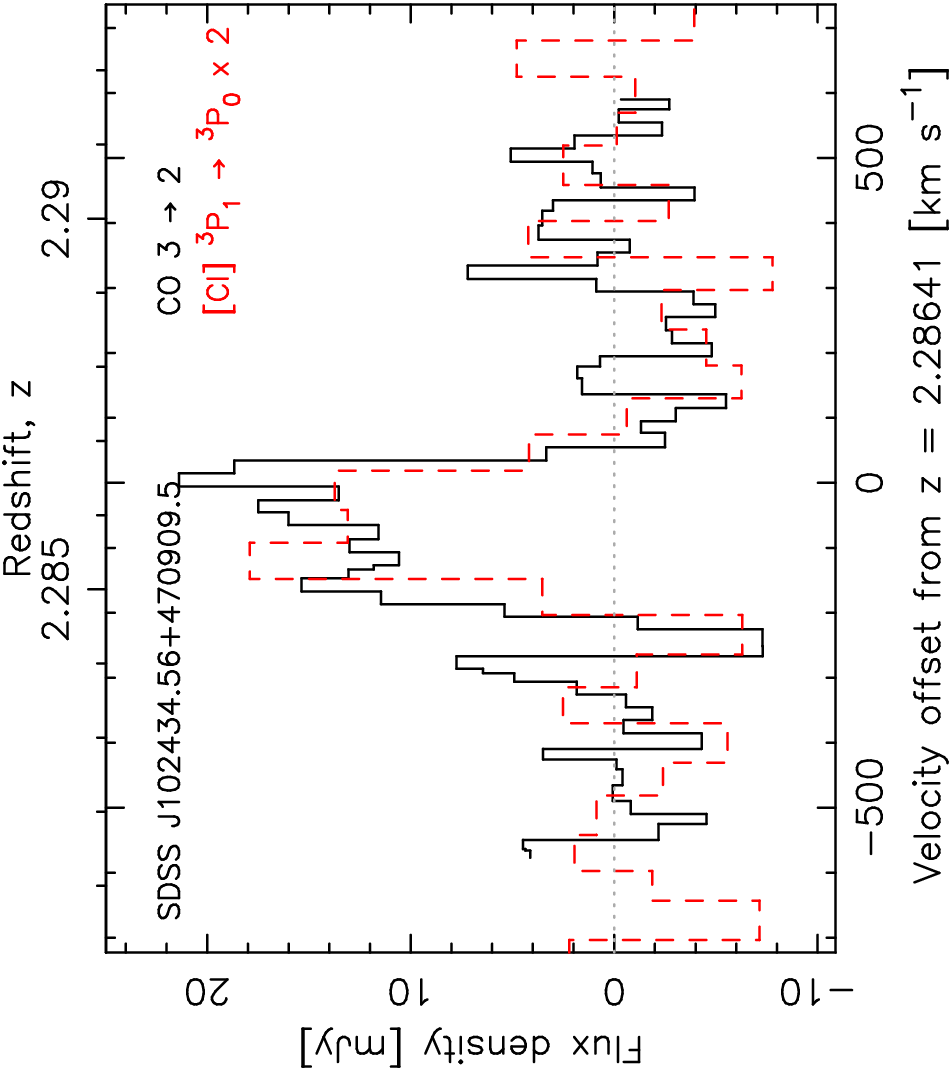}
\includegraphics{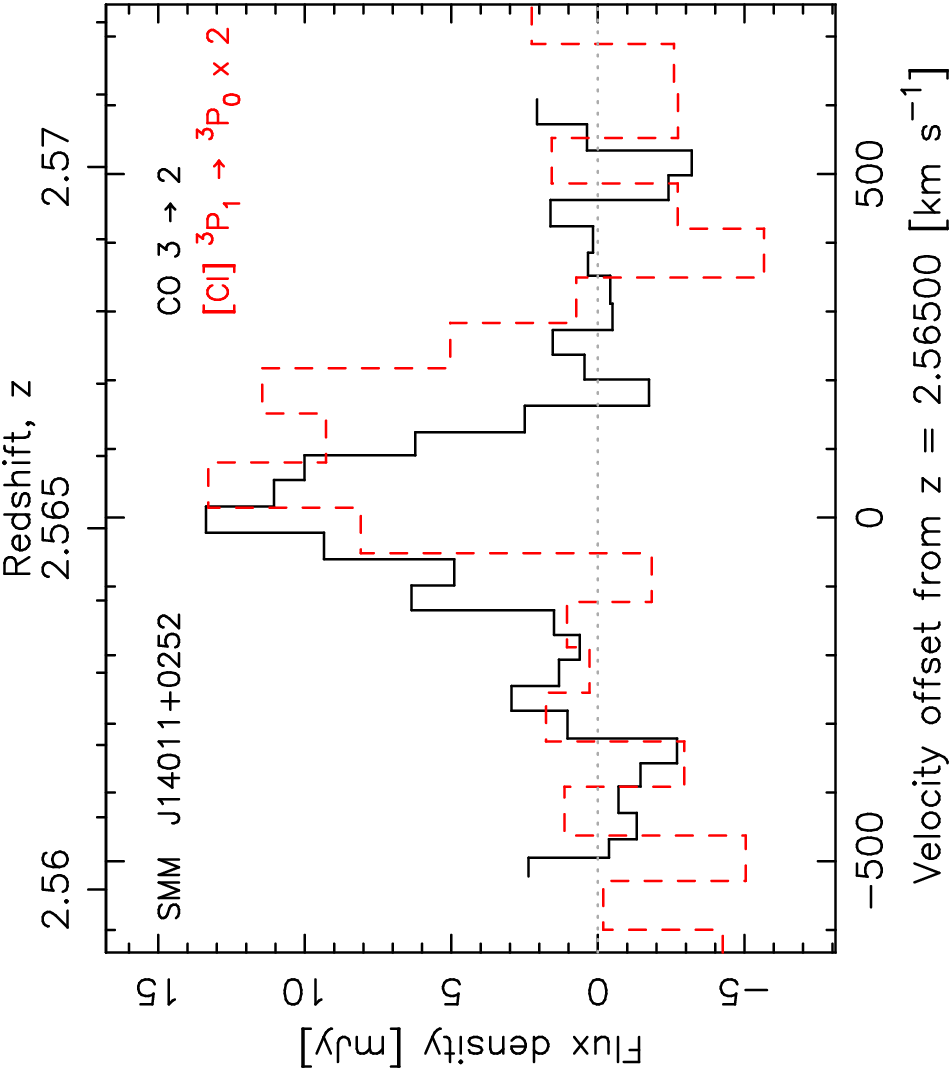}
\includegraphics{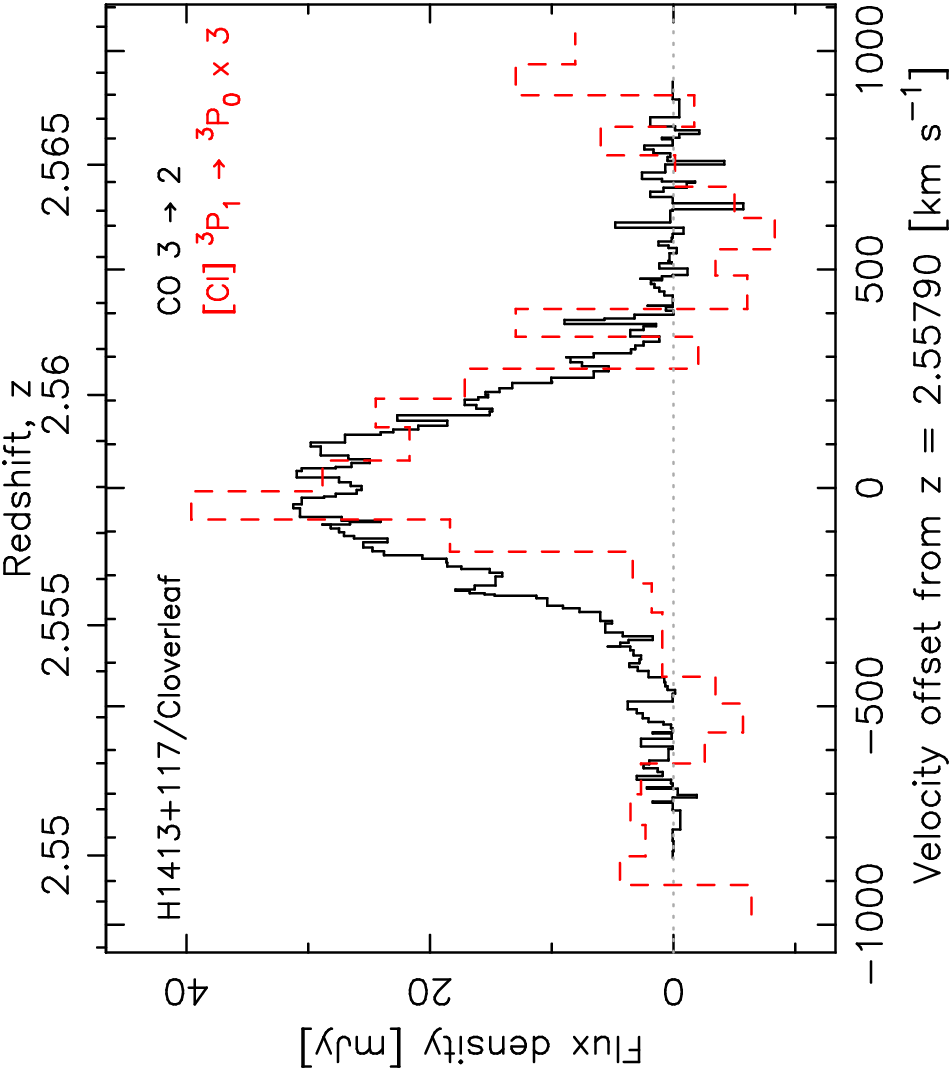}
\includegraphics{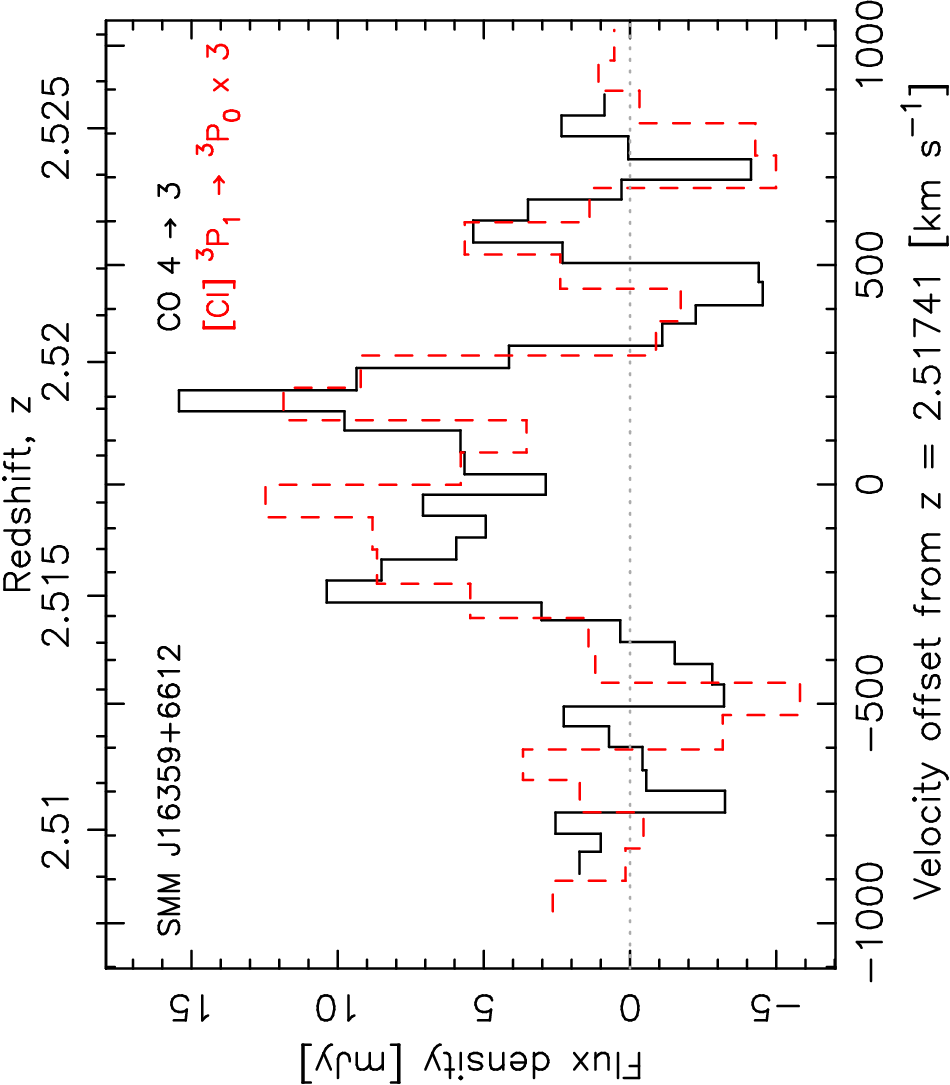}
\includegraphics{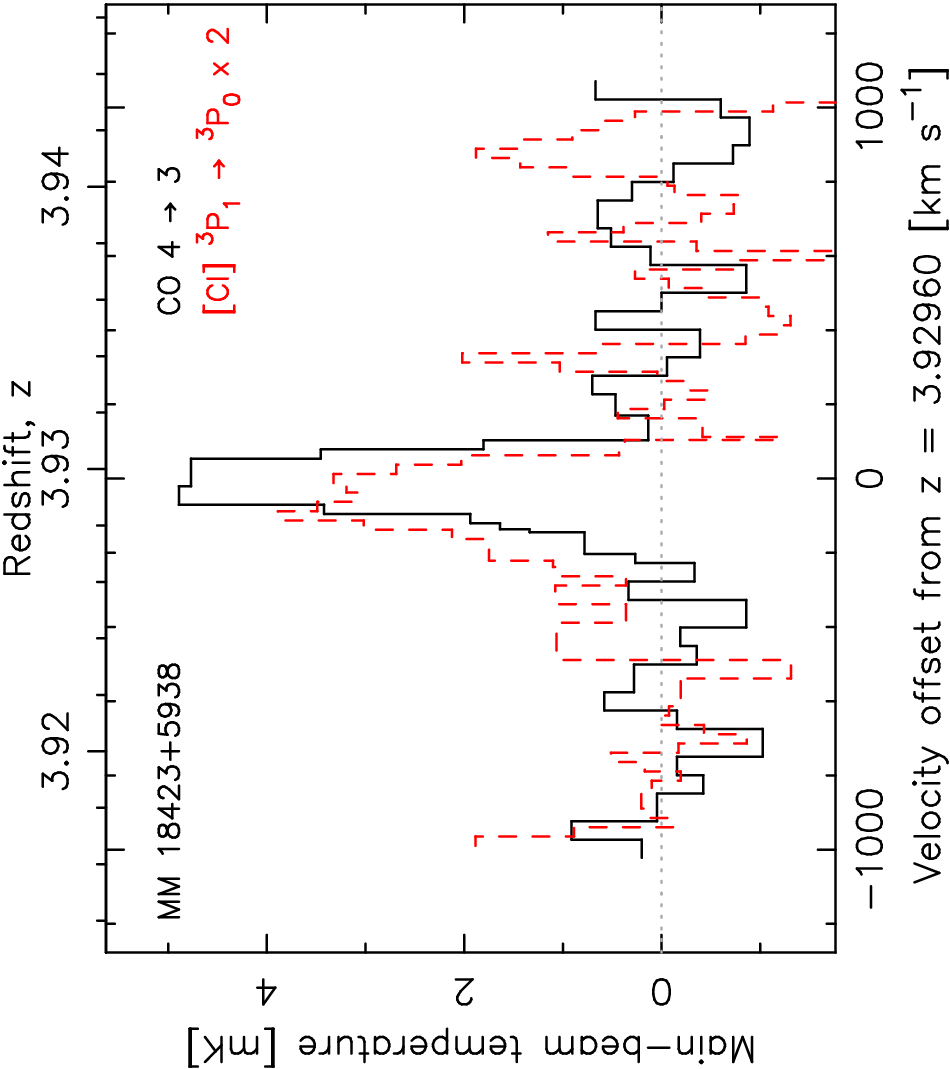}
\includegraphics{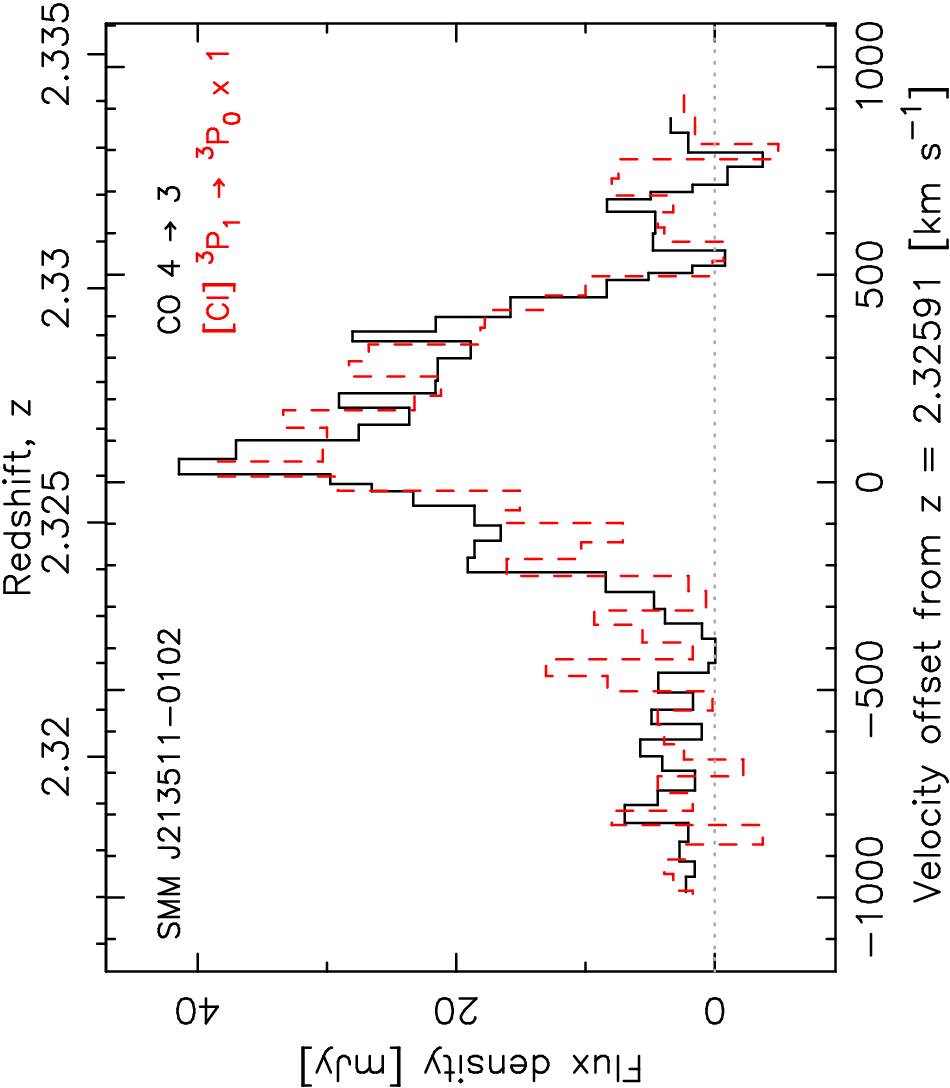}
\includegraphics{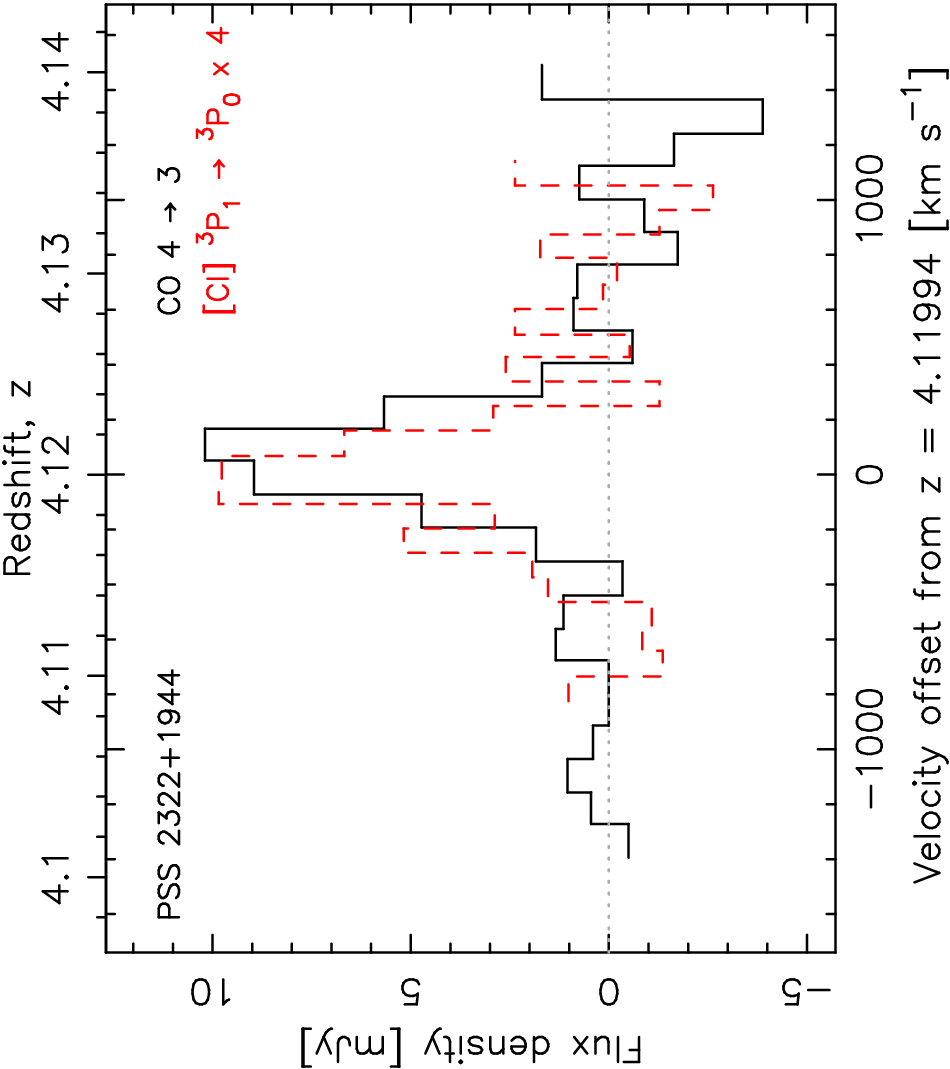}
\caption{The \carbi\ $^3{\rm P}_{1}\rightarrow\,^3{\rm P}_{0}$ spectra (broken lines) overlaid on the CO rotational transition 
used (full lines) for the published  high redshift \carbi\ detections, where the \carbi\ intensity is scaled by the value shown in the legend.
 All spectra are shown at the original spectral resolution and have
been re-sampled to the reference frequency of the CO transition ($\nu_{\rm ref}$ in Table \ref{fits}, thus showing the offset from this), except  in the
case of SMM\,J213511--0102 (the subject of a spectral scan), where both spectra are shown relative to the derived CO $1\rightarrow0$ redshift of $z =   2.32591$ \citep{dss+11}. 
Note that, for the sake of clarity, the spectra for MM\,18423+5938 have been truncated from the full $\pm2400$ \kms\ range \citep{lcs+10}.} 
\label{spectra}
\end{figure*}

In Fig. \ref{spectra} we show the $^3{\rm P}_{1}\rightarrow\,^3{\rm P}_{0}$ and rotational spectra of the sources in which both
\carbi\ and CO have been detected.\footnote{\citet{lsa+11}
claim a further two detections of the $^3{\rm  P}_{2}\rightarrow\,^3{\rm P}_{1}$ transition at $z = 1.786$ and $2.308$. However, 
these are so poorly resolved as to be indistinguishable from the CO $7\rightarrow6$ line and are thus of little use here.} 
These have been taken from the references listed in Table \ref{fits} and re-sampled in velocity according to the reference frequency
of the CO transition. Although the full-width half maxima (FWHMs) are similar between the \carbi\ and CO, it is seen that in some cases,
that there is a noticeable velocity offset between the profiles, which has also been noted by \citet{whdw03,pbc+04}.\footnote{There
are some differences in the FWHMs, particularly for  H\,1413+117 and APM\,08279+5255, the latter of which has the poorest \carbi\ spectral resolution
(150 \kms), although the former has the best (17 \kms). The fact that in some cases the CO spectrum
is wider than that of \carbi\ and vice versa (Table \ref{fits}), leads us to suspect that the differences may be due to the relatively low
quality of the spectra, which may also affect the velocity offsets (see Sect. \ref{sect:comp}).}

We analyse the spectra in a consistent manner by fitting a Gaussian to each using the {\sf gauss\_fit.pro} routine
in {\sc idl}.\footnote{Although several rotational transitions have been detected in some of the quasars
\citep{bmaa97,dnw+99,cod+02,bcn+03,wbc+03,tnc+06, awd+08,dss+11},
in order to minimise differences in the beam filling factors, where available, we use the CO transition ($J=4\rightarrow3$)
closest in frequency to the  \carbi\ line. Except in the case of H\,1413+117 (the Cloverleaf quasar), where the CO $3\rightarrow2$ \citep{whdw03}
spectrum is of higher quality than the $4\rightarrow3$ \citep{bmaa97}.} This uses a Lavenburg-Marquadt algorithm \citep{pftv89}, containing up to six terms;
 \[
F(x) = A_0 \exp\left\{-\frac{\left(x - A_1 \right)^2}{A_2^2} \right\} + A_3 + A_4 x + A_5 x^2,
\]
where $A_1$ is the mean of the fit, $A_2$ the standard deviation and $A_3,\,A_4$ and $A_5$ 
fit the continuum. Since there are no measured uncertainties for the data, {\sf gauss\_fit.pro} 
uses Poisson statistics to approximate data uncertainties, which is likely to overestimate the 
errors for the fits. 

Initially, fits were performed using all six terms in order to ensure complete removal of the continuum from the
data. Subsequent fits were then performed, adjusting the range of the data being fit, as well as the number of terms
used in the fit in order to ensure that the optimum fit was found (quantified by the lowest normalised $\chi^2$).  
Generally, we found that the fits with the smallest errors were those with $A_3, \, A_4$ and $A_5$ set to zero and,
when applied, these coefficients had uncertainties larger than their fitted values.

The results of the fits are summarised in Table \ref{fits}, where
the velocity offset ($\Delta v$) is defined relative to the reference frequency ($\nu_{\rm ref}$) for each spectrum, which was then
converted to a redshift ($z_{\rm fitted}$). The values for both the CO and \carbi\ spectra were then used to derive a fractional 
offset.\footnote{All of our analysis is done using the $^3{\rm P}_{1}\rightarrow\,^3{\rm P}_{0}$ transition and, although the $^3{\rm P}_{2}\rightarrow\,^3{\rm P}_{1}$ 
spectrum of \carbi\ has also been detected in a few of the sources -- RX\,J0911+0551 \citep{wwd+11},  J1148+5251 \citep{awd+08}, SMM \,J14011+0252 \citep{wwd+11},
H\,1413+117 \citep{whdw03}, MM\,18423+5938 \citep{lcs+10}, SMM\,J213511--0102 \citep{dss+11}, PSS\,2322+1944 \citep{wwd+11}, this is 
often blended with the CO $7\rightarrow6$ emission.}
\begin{table*}
\begin{minipage}{178mm}
\caption{Summary of the Gaussian fits to the \carbi\ $^3{\rm P}_{1}\rightarrow\,^3{\rm P}_{0}$ and the CO transitions, where we list all of the sources
in which a \carbi\ detection is claimed by \citet{wwd+11}, although spectra are currently available for only eight (Fig. \ref{spectra}).
These are in order of increasing redshift ($z_{\rm quoted}$, which is
derived from the centroid frequencies quoted in the reference). $\nu_{\rm ref}$ [GHz] is the frequency given in the reference from which the velocity offset,
$\Delta v$ [\kms], and fitted redshift, $z_{\rm fitted}$, are determined. FWHM [\kms] gives the full-width half maximum of the line as determined from the Gaussian fit (all single fits except SMM\,J16359+6612
for which two Gaussians were used and $\Delta v$ is the mean of the two fits).
The final column gives $\Delta F/F$, in units of $10^{-5}$, derived from the two values of $z_{\rm fitted}$.}
\label{fits}
\centering
\begin{tabular}{l  c  c c c r r c  r}
 \hline
\noalign{\smallskip}
Name & $z_{\rm quoted}$ & $\nu_{\rm ref}$ &  Ref & Trans & $\Delta v$ & FWHM &  $z_{\rm fitted}$ &  $\Delta F/F$ \\
\noalign{\smallskip}
\hline
SMM\,J123549+6215   & 2.202     & ---&  T06  & CO $3\rightarrow2$  & --- & --- & --- &---\\           
                                        & 2.202 &  ---   & W11& \carbi\ $^3{\rm P}_{1}\rightarrow\,^3{\rm P}_{0}$ & ---& --- &--- &---\\ 
 SDSS\,J102434.56         &       2.2854 &   105.220 & D95 & CO $3\rightarrow2$ & $-69\pm9$  &   $197\pm24$  & $2.28565\pm0.00010$ &  $10\pm4$\\
~~~+470909.5              & 2.2854  &   149.8024   &  W05a & \carbi\ $^3{\rm P}_{1}\rightarrow\,^3{\rm P}_{0}$ &  $-6.9\pm2.7$  & $157\pm31$ & $2.28532\pm0.00003$& ...\\
SMM\,J2135--0102  & 2.32591 &  138.563$^{\ast}$&  D11 & CO $4\rightarrow3$   & $0\pm13$& $550\pm30$& $2.32730\pm0.00014$ &  $-10\pm10$\\     
              & 2.326  &    147.901$^{\ast}$& D11 & \carbi\ $^3{\rm P}_{1}\rightarrow\,^3{\rm P}_{0}$ &  $0\pm16$ & $515\pm39$& $2.32764\pm 0.00018$ & ...\\ 
SMM\,J163658+4105 & 2.452 &  ---& T06   &  CO $3\rightarrow2$    &   --- & --- & ---& ---\\          
            & 2.452  &     ---& W11& \carbi\ $^3{\rm P}_{1}\rightarrow\,^3{\rm P}_{0}$ & ---& --- & ---& ---\\ 
SMM\,J16359+6612 & 2.51732&   131.074&  W05  & CO $4\rightarrow3$& $-6\pm21$& $410\pm110$&  $2.51733\pm0.00025$ & $-10\pm12$\\ 
                                   & 2.517 &   139.9229& W11 &\carbi\ $^3{\rm P}_{1}\rightarrow\,^3{\rm P}_{0}$ & $27\pm15$ & $430\pm90$ &  $2.51769\pm 0.00018$ & ...\\ 
H\,1413+117    &      2.55784 &   97.191  & W03 & CO $3\rightarrow2$ & $-14\pm2$  &  $421\pm5$ &  $2.55774\pm0.00002$ & $-4\pm8$\\
 ~~~   (Cloverleaf)              & 2.5578  &  138.3313   &  W05a &\carbi\ $^3{\rm P}_{1}\rightarrow\,^3{\rm P}_{0}$ & $4\pm22$ &  $329\pm52$ & $2.55789\pm0.00026$ & ...\\
SMM \,J14011+0252     &      2.5653 &   96.9975  & F99 & CO $3\rightarrow2$ &  $27\pm7$ & $189\pm16$  &   $2.56532\pm0.00008$ &$-14\pm8$\\
                  & 2.5653 &    138.0457  &  W05a & \carbi\ $^3{\rm P}_{1}\rightarrow\,^3{\rm P}_{0}$ & $51\pm18$ &  $251\pm40$ &  $2.56581\pm0.00021$ & ...\\
RX\,J0911+0551 &   2.796&   91.088& H04& CO $3\rightarrow2$ & ---& --- & ---& ---\\ 
     & 2.796       &  ---& W11 &  \carbi\ $^3{\rm P}_{1}\rightarrow\,^3{\rm P}_{0}$ &  ---& --- & ---& ---\\ 
SMM\,J02399--0136 &    2.8076   &   90.81 & G03 &  CO $3\rightarrow2$ & ---&  ---& ---&  ---\\ 
                                   & 2.808        & --- & W11 & \carbi\ $^3{\rm P}_{1}\rightarrow\,^3{\rm P}_{0}$ & ---&  ---& ---& ---\\   
APM\,08279+5255    & 3.9114   &93.867   &D99 &    CO $4\rightarrow3$ &   $-20\pm12$ &   $571\pm28$ & $3.91131\pm0.00019$ &  $-36\pm9$\\
                  & 3.913 &  100.216 &  W06 &  \carbi\ $^3{\rm P}_{1}\rightarrow\,^3{\rm P}_{0}$   &  $128\pm17$ &  $414\pm49$ & $3.91309\pm0.00027$ & ...\\  
MM\,18423+5938  & 3.92960 &  93.5249$^{\dagger}$ & L10& CO $4\rightarrow3$  & $-30\pm5$& $162\pm12$&  $3.92911\pm0.00009$ & $18\pm4$\\ 
                                   & 3.930 &  99.8378$^{\dagger}$& L10 &\carbi\ $^3{\rm P}_{1}\rightarrow\,^3{\rm P}_{0}$ &  $-83\pm6$& $230\pm14$ &  $3.92825\pm0.00010$ &...\\ 
 PSS\,2322+1944       &   4.1199  &   90.048   &C02 & CO $4\rightarrow3$ &  $ 3\pm17$ &   $361\pm42$ &  $4.11999\pm0.00029$ & $17\pm13$\\
                 & 4.1199  &     96.127  & P04 & \carbi\ $^3{\rm P}_{1}\rightarrow\,^3{\rm P}_{0}$ & $-46\pm22$ &   $277\pm59$& $4.11911\pm0.00037$ & ...\\ 
  SDSS\,J1148+5251 & 6.4189  & 93.204& B03 &  CO $6\rightarrow5$ & ---& ---& ---& ---\\ 
   & 6.419 &       --- & R09     & \carbi\  $^3{\rm P}_{2}\rightarrow\,^3{\rm P}_{1}$    &   ---&  ---& ---& ---\\
\noalign{\smallskip}
\hline
\end{tabular}
{\flushleft 
$^{\ast}$The absolute frequency and $^{\dagger}$the centroid frequency of the line from a spectral scan.\\ 
References:  D95 -- \citet{dsr95},  D99 -- \citet{dnw+99}, F99 -- \citet{fis+99}, C02 -- \citet{cod+02}, G03 -- \citet{gbt+03}, B03 -- \citet{bcn+03}, W03 -- \citet{whdw03},  
H04 -- \citet{hsy+04},
P04 -- \citet{pbc+04}, 
 W05a -- \citet{wdhw05}, W05b -- \citet{wdwh05}, T06 -- \citet{tnc+06}, W06 --  \citet{wwn+06}, R09 --  \citet{rwcl09} [according to \citealt{wwd+11}], L10 -- \citet{lcs+10}, D11 -- \citet{dss+11}, W11 -- \citet{wwd+11}.}
\end{minipage}
\end{table*}

\subsection{The comparison of rotational and fine structure transitions}
\label{sect:comp}

\begin{figure*}
 \centering
\includegraphics[angle=270,scale=0.70]{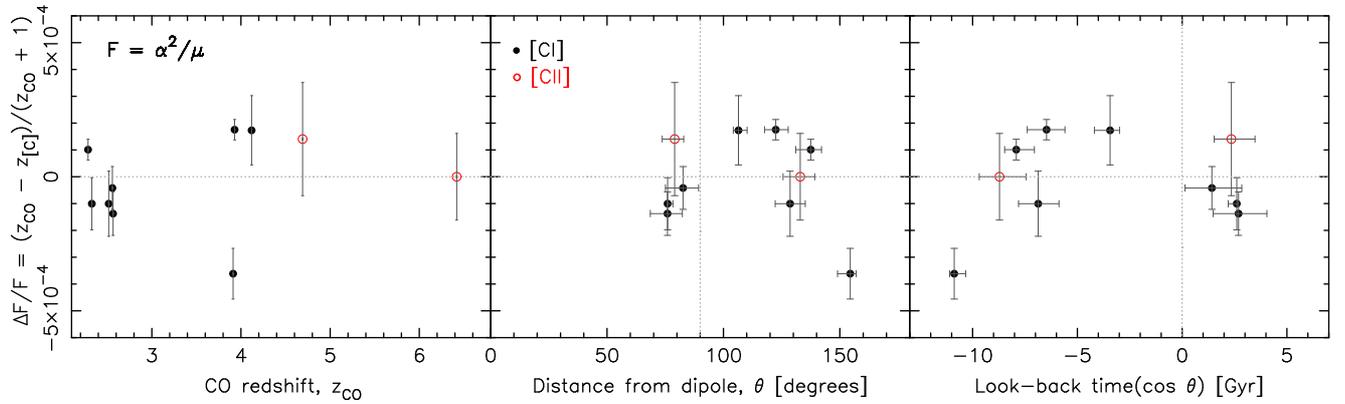}
\caption{$\Delta F/F$ derived from the velocity offsets in Table \ref{fits} versus the redshift (left), distance from the dipole (middle) and the
product of the look-back time with the cosine of the distance from the dipole (right), cf. \citet{wkm+10}. The filled black markers
show the eight \carbi\ sources and the unfilled colour markers the two \carbii\ sources (see \citealt{lrk+08}).}
\label{CvCO}
\end{figure*}

The rotational lines of CO are proportional to $\mu$, while the fine structure lines of neutral or ionised carbon (represented by [C])
are proportional to $\alpha^{2}$. Comparison of the redshifts of each species measures
\[
\frac{z_{\rm CO} - z_{\rm [C]}}{z_{\rm CO} +1} = \frac{\Delta F}{F},  ~{\rm where}~F=\frac{\alpha^2}{\mu}.
\]
Showing the values for $\Delta F/F$ derived from the \carbi\ lines, along with those from the \carbii\ \citep{lrk+08}, 
we find that $\alpha^2/\mu$ exhibits little evolution with redshift (Fig.~\ref{CvCO}, left), giving a mean value of 
$<\Delta F/F> = (-3.6\pm8.5)\times10^{-5}$, for the \carbi\ systems only, and $(-1.5\pm11)\times10^{-5}$,
including the \carbii\ systems, over $z = 2.3 - 4.1$ and $z = 2.3 - 6.4$, respectively.
Thus, the change in $\alpha^2/\mu$ is zero, within the observational ``noise''.

Referring to Fig.~\ref{CvCO}, however, there {\em may} a decrease in $\Delta F/F$ with distance from the dipole (middle panel) and an increase 
with $r\,\cos\theta$ (right panel) evident, as has been found for  $\Delta \alpha/\alpha$ \citep{wkm+10}.
The observed range in $\Delta F/F$ spans a change of $\sim40$ times that of $\Delta \alpha/\alpha$
($\lapp4\times10^{-4}$, cf. $\lapp1\times10^{-5}$) and interestingly, some theoretical models predict temporal variations of 
$|\Delta\mu/\mu|\sim 40\,|\Delta \alpha/\alpha|$ \citep{uza03}.  However, such models are tentatively ruled out by recent
observations, which find $|\Delta \mu/\mu| \lapp 10^{-6}$ at $z=0.89$ \citep{hmm+09,mbg+11} and $\lapp 3\times10^{-6}$ at
$z=2.81$  \citep{kmuw11}.

\begin{figure}
 \centering
\includegraphics[angle=270,scale=0.78]{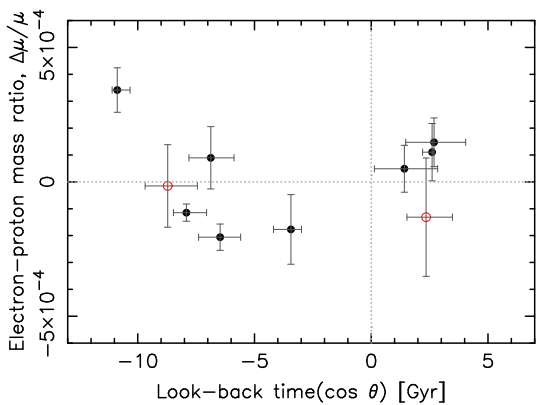}
\caption{The fractional change in the electron-proton mass ratio  versus the product of the look-back time and the cosine of the distance from the dipole,
upon the subtraction of $2\Delta \alpha/\alpha$  according to the best-fit dipole \citep{wkm+10}. The symbols are as per Fig. \ref{CvCO}.}
\label{CvCO-alpha}
\end{figure}

Removing the dependence of the fine structure constant on the dipole [$\Delta \alpha/\alpha = (1.1\pm0.2)\times10^{-6}\, r\, \cos\theta  - 
(1.9\pm0.8)\times10^{-6}$, where $r$ is the look-back time], 
via, $\Delta\mu/\mu = 2\,\Delta\alpha/\alpha -  \Delta F/F,$
gives the distribution shown in Fig.~\ref{CvCO-alpha}.
However, as per those in Fig. \ref{CvCO}, any apparent trend is due to the inclusion of
APM\,08279+5255 (at -10.88 Gyr and  $\Delta F/F = 3.4\times10^{-4}$), which  has the weakest \carbi\ spectrum (Fig.~\ref{spectra}), while exhibiting a large offset from the CO redshift
(in our fit as well as by \citealt{wwn+06}). 
Indeed,
several of the \carbi\ detection profiles are over very few channels, presumably to increase the signal-to-noise ratio of the
spectrum.

\begin{figure}
 \centering
\includegraphics[angle=270,scale=0.67]{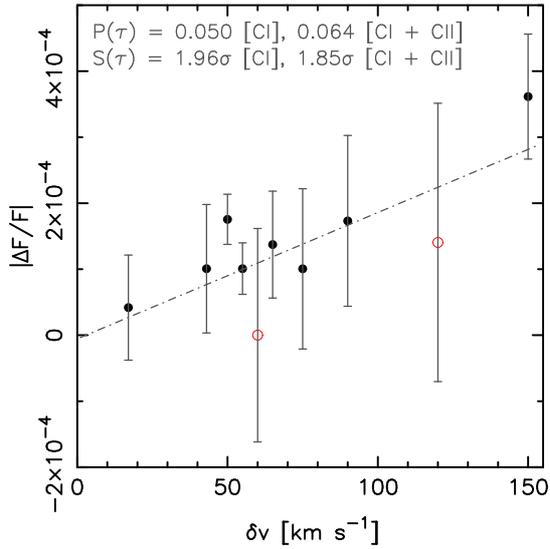}
\caption{The absolute value of $\Delta F/F$ versus the spectral resolution of the \carbi\ and \carbii\ detection. The symbols are as per Fig.
\ref{CvCO} and the line shows the least-squares fit to all of the points. $P(\tau)$ shows the Kendall's two-sided probability 
of the distribution occuring by chance and $S(\tau)$ the corresponding significance for the \carbi\ and \carbii spectra.}
\label{F-spec}
\end{figure}

In order to quantify the quality of the carbon spectra, in Fig.~\ref{F-spec} we show the derived value of $\Delta F/F$
versus the spectral resolution, $\delta v$, from which we see a distinct correlation.
It is surprising that this has a $\leq6$\% probability of occuring by chance, when ideally $\Delta F/F$ should
exhibit no dependence on $\delta v$. This is strong evidence that any large values of $\Delta F/F$ in the present data
are the result of low quality spectra. In an attempt to compensate for this, 
if we weight each value of $\Delta F/F$ by $1/\delta v$, thus having the lower resolution spectra making a smaller
contribution to the mean, we obtain $<\Delta F/F>_{\rm weighted} = (-2.0\pm8.0)\times10^{-5}$ (\carbi\ only) and
$(-1.2\pm9.3)\times10^{-5}$ (\carbi\ and \carbii), i.e. closer to zero variation.

This interpretation of the measured values of $F$ would explain why the temporal variation of $F$ as indicated by APM\,08279+5255 is at odds with
other sources at similar redshifts, which all exhibit positive values of $\Delta F/F$ (Fig. \ref{CvCO}, left panel). As discussed
above,  such a discrepancy may be expected due to spatial variation of the constants, were the sources in widely separated regions of the sky.
However,  MM\,18423+5938 and  PSS\,2322+1944 are at 122 and 106 degrees from the dipole, respectively, whereas
APM\,08279+5255  is at 154 degrees (Fig.~\ref{CvCO}, middle panel), thus being in the same hemisphere as the other two sources. This 
supports the suggestion
that the derived offset of the \carbi\ redshift in APM\,08279+5255 does not accurately trace $\Delta F/F$.

Apparent offsets may also arise from calibration errors, which have dominated previous measurements of the constants using neutral
carbon: The comparison of the ultra-violet neutral carbon resonance lines with the radio-band \HI\ 21-cm hyperfine transition yield $\Delta
X/X$, where $X \equiv g_{\rm p}\,\alpha^2\,\mu$, with $g_{\rm p}$ being the proton gyromagnetic factor. The three sources in which both
species have been detected (all in absorption) have yielded $\Delta X/X = (7\pm11)\times10^{-6}$ at $z=1.78$ \citep{cs95} and 
$(7\pm7)\times10^{-6}$ at $z=1.36 - 1.56$ \citep{kpec10}. The relatively large uncertainties are systematic, arising from the absolute
wavelength calibration of the optical spectrum. This is an issue which we avoid through the comparison of redshifted submillimetre
transitions, since the frequency calibration of  radio and millimetre-wave instruments is phase-locked to an atomic clock frequency standard. Although
due to the low spectral resolution of the \carbi\ detections and the fact that the spectra are emission profiles from entire galaxies, our absolute uncertainties are much larger.

Lastly, it is entirely possible that the relative velocity shifts between the species are intrinsic, although as noted above, the values of
$\Delta F/F$ are consistent with observational noise, with most of the error bars in Fig. \ref{CvCO} extending to $\Delta F/F\approx0$.
Furthermore the FWHMs of the CO and \carbi\ emission profiles are very similar and the fact that CO is wider than \carbi\ in only about half of
the cases (5 out of 8, although the uncertainties overlap), may indicate that the FWHMs could be the same with apparent differences introduced by the low spectral resolution of the
data.\footnote{The mean FWHM of the eight CO spectra is $360 \,(\sigma = 150)$ \kms, cf. $330 \,(\sigma = 110)$ \kms\ for the \carbi\ spectra,
based upon the values in Table \ref{fits}.} On this issue, it is also possible that systematics may be introduced by unresolved components in the data, again reinforcing the need for
higher resolution data. Such data may also be used to ascertain the validity of fitting Gaussians to the profiles, which we believe is a
reasonable assumption for the distribution of gas in a galaxy (with the number of Gaussians required depending on the orientation of the
galaxy, e.g. \citealt{cur99p}). However, the fact that these fits can yield ``accuracies'' to well within a single channel, gives the relatively
small error bars for MM\,18423+5938 (at $z=3.9$ and $\Delta F/F = 1.8\times10^{-4}$, Fig.~\ref{CvCO}). Specifically, $\pm6$ \kms\ from our
fit, cf. the $\pm24$ \kms\ (half a channel) quoted by \citet{lcs+10}. Thus, an artificially small uncertainty could arise
since, of
all the \carbi\ spectra, this is the most Gaussian shaped (Fig. \ref{spectra}), when in reality the resolved emission may be 
too complex to be best fit by a single Gaussian. Again this is an issue which can only be addressed with higher 
resolution carbon spectra.

To surmise, based upon the current eight high redshift \carbi\ and two \carbii\ detections, the data appear to be
dominated by scatter. We have shown that this is, at least in part, due to the low spectral resolutions of the 
carbon detections. 
Since current instruments are generally capable of much finer resolutions, we suspect
that such coarse values have been used in order to tease as strong a detection as possible from the data,
leading to the scatter in $\Delta F/F$.
Therefore in order to make meaningful measurements of the constants, spectra with much higher signal-to-noise
ratios are required. 

\section{Improving the data}
\subsection{Prospects with current instruments}
\label{pci}

\begin{figure}
 \centering
\includegraphics[angle=270,scale=0.67]{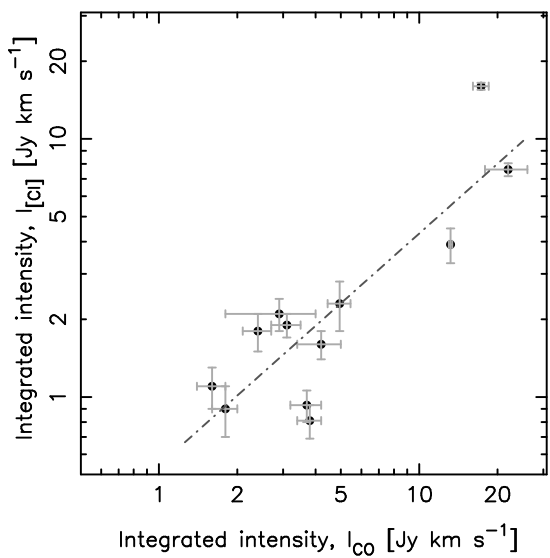}
\caption{The velocity integrated intensity of the \carbi\ $^3{\rm P}_{1}\rightarrow\,^3{\rm P}_{0}$ line versus that of the CO
transition quoted in Table \ref{fits}. The main-beam temperature of MM\,18423+5938 is converted to Jy using a conversion of
5.0 Jy K$^{-1}$ \citep{lcs+10}. The line shows the least-squares fit to all of the points.}
\label{intensity}
    \end{figure}

\begin{figure*}
 \centering
\includegraphics[angle=270,scale=0.60]{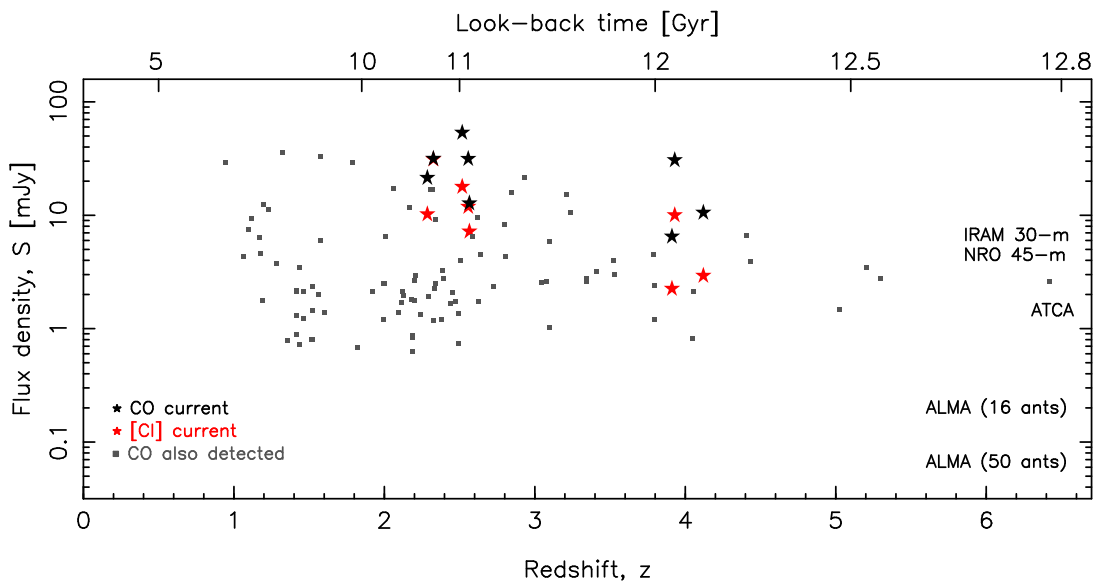}
\caption{The flux density ($S = I/{\rm FWHM}$) of the CO and \carbi\ emitters versus redshift. The black stars show the current redshifted CO emitters
which also exhibit \carbi\ emission (designated by coloured stars).
The small squares show all of the other published $z\geq1$ CO emitters. 
The telescope labels show the expected
r.m.s. noise level per each 50 \kms\ channel after 10 hours of observation with the corresponding telescope at 100 GHz (see main text).
The look-back time is calculated using $H_{0}=71$~km~s$^{-1}$~Mpc$^{-1}$, $\Omega_{\rm
    matter}=0.27$ and $\Omega_{\Lambda}=0.73$ (\citealt{svp+03}), which we use throughout the paper.}
\label{flux-z}
 \end{figure*}

Of the known high redshift CO emitters\footnote{A total of 103 published thus far 
(\citealt{dsr95,oyn+96,syw+97,fis+98,fis+99,prv+00,cod+02,dnm+03,gip03,wbc+03,whdw03,hsy+04,ddn+05,
kes+05,kns+05,sv05,itn+06,tnc+06,csn+07,mnb+07,wmr07,ccn+09,dds+09,wid+09,bct+10,dbw+10,lcs+10,rcc+10,tgn+10,ytf+10,dss+11,enf+11,
fhb+11,lsa+11,rie11a} and references therein).}, 
most have the $^3{\rm P}_{1}\rightarrow\,^3{\rm P}_{0}$ transition of \carbi\ redshifted into current millimetre bands
and are therefore potentially detectable with current instruments, expanding the dataset by an order of magnitude. 
In order to determine the expected emission strengths
of the \carbi, from the current detections and those listed in \citet{wwd+11}, we find a mean of 
$I_{\rm [CI]}\approx 0.5\pm0.2\,\times \,I_{\rm CO}$ (Fig.~\ref{intensity}), cf. the ratios of $\approx0.1$ within the Galaxy \citep{osh+01,iot+02},
$\approx0.5$ in M82 \citep{wec+94,sgh+97} and $\approx0.2$ in other near-by galaxies \citep{gp00}.
Since the FWHM of the \carbi\ profile is close to that of the CO in the detected cases (Fig. \ref{spectra}),
we can replace the integrated intensities with the line fluxes,
in order to estimate the relative telescope sensitivities (Fig. \ref{flux-z}).

In  the figure the expected r.m.s. noise levels achievable by the telescopes are derived for  a spectral resolution of 50 \kms\ (17 MHz at 100 GHz, typical of the current \carbi\ detections),
after 10 hours on source. 
We have used the specified system temperature of $T_{\rm sys} = 250$~K for each, which gives an r.m.s of 6 mJy for the  Institut de Radioastronomie Millim\'{e}trique (IRAM) 30-m, 4 mJy for the 
Nobeyama Radio Observatory (NRO) 45-m
and 1.3 mJy for the Australia Telescope Compact Array (ATCA). 
These sensitivities do indicate that more detections are to be expected over all redshift ranges with current instruments.

As discussed above, however, in order to measure velocity offsets of high enough quality to measure variation in the constants, the
spectral resolution must generally be much finer than for the current \carbi\ detections, 
requiring significantly lower r.m.s. noise levels than shown in Fig. \ref{flux-z}.  Furthermore, even with
sufficient spectral resolution, current spatial resolutions may be insufficient: The finest of the current carbon
detections subtend (unlensed) linear extents in excess of 20 kpc (HPBWs$\,\approx2"$,
\citealt{dsr95,dnw+99,whdw03}), although $\gapp100$ kpc is more common.  Since, we
ideally require a comparison of redshifts from a single complex, namely a giant molecular cloud (GMC), resolutions of
$\lapp100$ pc are required, which even the finest spatial resolutions currently available cannot resolve:
For example, with a maximum 6-km baseline, the ATCA is capable of a $0.3"$ resolution at 100 GHz,
which gives linear extents of between 2.2 kpc (at $z=3.7$) and $1.9$ kpc (at $z=4.9$, the \carbi\ redshift range
observable with the ATCA 3-mm band).

\subsection{Prospects with ALMA}

On the basis of their sensitivity and spatial resolution, it is clear the current instruments are of limited use in
using carbon to measure cosmological variation of the constants. With the ALMA {\em
  Early Science} configuration ($16\times12$-m antenna), the maximum 400 metre baseline gives a beam of $\gapp2"$ at 100~GHz, i.e. a linear
extent of $\gapp1$~kpc at $z\sim4$ and, after 10 hours on source, an r.m.s. noise level of 0.19 mJy
per each 50 \kms channel is expected (Fig.~\ref{flux-z}).
For the {\em Full Array} configuration ($\geq50\times12$-m antennas), $\leq0.06$ mJy per 50 \kms\ channel
will be reached after 10 hours on source. The finest beam-width
will be $\approx0.05 - 0.04"$ over 84--116 GHz (band-3), which corresponds to linear extent of
330 pc at $z=4.9$ and 300 pc at $z=3.2$. This level of resolution is getting
close to the $\lapp100$ pc required to resolve individual GMCs and, in the case
where the galaxy is gravitationally lensed by an intervening galaxy, a magnification of only $\gapp3$ is required to 
achieve sub-100 pc resolution.

\begin{figure*}
 \centering
\includegraphics[angle=270,scale=0.60]{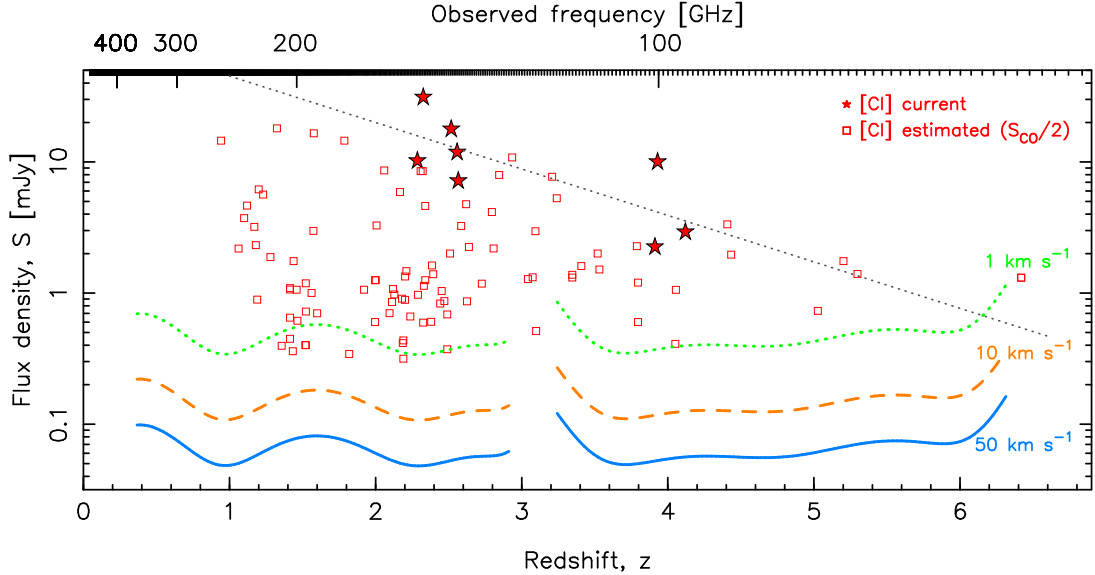}
\caption{As per Fig. \ref{flux-z} but showing the  expected flux density of the \carbi\ emission based upon 
$S_{\rm CO}\approx 2\times S_{\tiny\rm \carbi}$  (Fig. \ref{intensity}). The 
dotted, broken and full traces show the r.m.s. sensitivity of the full ALMA array per each 1, 10 and 50 \kms\ channel
for the conditions used above (10 hours on source  at a declination of $\delta = -30$\dg\ and
a vapour column density of 1.262 mm). The lower redshift trace spans the range given by the band-7, 6, 5 and 4 receivers (370 -- 125 GHz)
and the higher redshift spans that given by the band-3 and 2 receivers (116 -- 67 GHz). 
The frequency scale is derived for the redshift of the \carbi\ $^3{\rm P}_{1}\rightarrow\,^3{\rm P}_{0}$ transition.
The stars show the known high redshifted \carbi\ emitters with the dotted line showing the least-squares fit to these.}
\label{flux-alma}
 \end{figure*}

With ALMA, we are
thus in the realm of Galactic resolution millimetre astronomy at very large look-back times. As such,
the spectral resolutions used for the eight \carbi\ detection experiments (Sect. \ref{sect:comp}) are insufficient in
the study of individual GMCs, as well as the application of these in measuring variation in the constants,
where, for a ratio of line frequencies of a few parts in $\sim10^5$, we
require spectral resolutions of better than 10 \kms. 
Using the ALMA sensitivity calculator\footnote{http://almascience.eso.org/call-for-proposals/sensitivity-calculator}, at 
the above (50 \kms) and finer (10 \& 1 \kms) spectral resolutions, we show the r.m.s. noise levels expected from the
{\em Full Array} configuration after 10 hours on source (Fig. \ref{flux-alma}).
From this, we see that the currently detected \carbi\ emitters are likely to be the most luminous of this sample, with the estimated fluxes
of those yet to be searched
generally lying below the least-squares fit described by these. This suggests that any further detections with
current instruments are unlikely to be of higher quality than those in  Fig.~\ref{spectra}, while
the ALMA {\em Full Array}
will be capable of detecting \carbi\ in all\footnote{Apart from six, but not on the basis of sensitivity, but that 
the redshifted \carbi\ $^3{\rm P}_{1}\rightarrow\,^3{\rm P}_{0}$ transitions falls between the  band-4--7 and band-2--3 receivers at 116--125 GHz ($z=2.94 - 3.24$).}
of these at a spectral resolution of 10 \kms\ or better.

As seen from Fig. \ref{flux-alma}, most of the redshift space is covered, thus making the total number of sources expected to be detected
close to the number quoted above. That is, based upon the current CO detections {\em only},  the {\em Full Array} configuration is expected to 
detect $\sim100$ galaxies in \carbi\ at better than 10 \kms\ spectral resolution. With regard to individual complexes, GMC numbers
in external galaxies can range from a few (e.g. 6 in M31, \citealt{svwd08}) to over a hundred (148 in M33, \citealt{eprb03}), where observations
are of sufficient sensitivity. Therefore, as an order of magnitude estimate, we can expect the detection of $\gapp1000$  \carbi\ emitters to
very large look-back times, a much larger sample with which to measure the constants than available in the optical regime.

\section{Conclusions}

By fitting the carbon and carbon monoxide  emission spectra,
we have determined the relative velocity offsets between these species in the
eight redshifted sources in which the $^3{\rm P}_{1}\rightarrow\,^3{\rm P}_{0}$ transition of neutral carbon
has been detected. Such comparison provides a measure of the fundamental constants,
specifically $F\equiv\alpha^2/\mu$, where \AL\ is the fine structure constant and $\mu$ is the 
electron-proton mass ratio. Being a comparison of fine structure (\carbi) and rotational lines (CO), this gives the same combination
of constants as yielded by the comparison of ionised carbon (\carbii) with CO \citep{lrk+08}, although it has the distinct advantage
that the neutral carbon may be spatially coincident with the CO, within molecular clouds, whereas the ionised carbon is
located between clouds.

We find $<\Delta F/F> = (-3.6\pm8.5)\times10^{-5}$ for the eight \carbi\ systems and 
$<\Delta F/F> = (-1.5\pm11)\times10^{-5}$, when the two \carbii\ systems are included, the former
of which we believe to be the most reliable due to the segregation of \carbii\ and CO. Both results are consistent with
a zero variation over the redshift range $z = 2.3 - 6.4$ (look-back times of $10.8-12.8$ Gyr), although $\Delta F/F$ does appear
to vary in the same direction as the spatial dipole defined by the variation in  
\AL\ \citep{wkm+10}. This, however, is dominated
by the lowest quality \carbi\ spectrum, which has a channel spacing of 150~\kms, giving the largest velocity offset
between the \carbi\ and CO spectra and with the removal of this source  no trend is apparent.

Quantifying the quality of the spectra, we find that large values of $\Delta F/F$ are favoured by the lower quality carbon spectra, with
$|\Delta F/F| \rightarrow 0$ as $\delta v  \rightarrow 0$, where $\delta v$ is the spectral resolution of the carbon detection. 
Although the sample is small, this is strong evidence that the non-zero values of $\Delta F/F$ are the result of low quality
spectra.  Conversely, however, the zero variation of constants found from the ionised carbon lines are also unreliable \citep{lrk+08} by this
reasoning. Weighting the contribution of each measurement of $\Delta F/F$ by $1/\delta v $, gives 
$<\Delta F/F>_{\rm weighted} = (-2.0\pm8.0)\times10^{-5}$ for \carbi\ only and $(-1.2\pm9.3)\times10^{-5}$
when the two \carbii\ systems are included.

Such low spectral resolutions are the result of maximising the signal-to-noise ratio in order to yield the
detection. Applying the relation $I_{\rm CO}\approx 2\times I_{\rm [CI]}$ to the $z\gapp1$ sources, indicates that the
current \carbi\ detections would be among the most luminous of the sources already
detected
in CO. That is, even if current instruments were to detect \carbi\ emission in the remainder of the CO emitters, the spectra are unlikely to be of sufficient quality to provide meaningful
measurements of the constants. Furthermore, from the spatial resolution and spectral profiles of the
current detections, it is clear that we are detecting emission over much of the galaxy, when ideally
we require comparison of the \carbi\ and CO lines in a single complex. As such, the
best spatial resolution of $\gapp2$ kpc, currently available in the 3-mm band ($z_{\rm [CI]}=3.7-4.9$),
is incapable of resolving these complexes.

The sensitivity and spectral resolution of ALMA is therefore required in order to fully utilise the
use of neutral carbon and carbon monoxide in measuring the constants:
\begin{itemize}
\item After 10 hours on source, the {\em Early Science} configuration
($16\times12$-m antennas) is expected to reach an r.m.s. noise level of $\approx0.2$ mJy per each 50 \kms channel.
However:
           \begin{itemize}
             \item Although sufficient to detect \carbi\ in all of the currently known $z\gapp1$ CO
emitters (Fig, \ref{flux-z}, cf. Fig. \ref{flux-alma}), it is clear from this work that spectral resolutions of $\approx50$ \kms\
 are insufficient. 
         \item Furthermore, the  maximum linear resolution of $\gapp1$~kpc, at $z\sim4$, is still too coarse to resolve individual GMCs.
         \end{itemize}
\item  After 10 hours on source, the {\em Full Array} configuration
($\geq50\times12$-m antennas) is expected to reach a similar  r.m.s. noise level per each 10 \kms channel (Fig. \ref{flux-alma}).
          \begin{itemize}
       \item This is sufficient to detect all of the current CO emitters in \carbi\ which fall into bands-3 to 7 ($z = 0.33 - 6.36$) at
         the spectral resolutions required.
         \item The maximum angular resolution of $\approx0.04"$ at 116 GHz, gives a spatial resolution of $\approx300$ pc at 
           $z = 3.2 - 4.9$, which is still insufficient to resolve a $\lapp100$ pc GMC. However, the majority of distant galaxies in the 
           submillimetre bands may be subject to magnification by gravitational lensing \citep{bla96}, as are all of the current \carbi\ detections (\citealt{orm03,gkv05,iss+10}
           and references therein).\footnote{No such data yet exists for MM 18423+5938, which has yet to make it onto the NASA/IPAC Extragalactic Database, although
           the discovery paper \citep{lwb+09} does note multiple components.} Lens magnifications of $\gapp40$ may be expected \citep{bmm99}, although magnifications of only
           $\gapp3$ are required for ALMA to achieve $\lapp100$ pc resolution at these redshifts.
  
         \end{itemize}
\end{itemize}
Detecting \carbi\ in just each of the current $z\gapp1$ CO emitters would bring the number of systems
to $\sim100$, which is getting close to the large samples of optical spectra
which indicate a space-time variation in \AL\ \citep{mwf03,wkm+10}. Being able to resolve individual GMCs
in these  would take this number to $\gapp1000$ individual systems, based upon the currently detected  CO
emitters {\em alone}. This will yield an invaluable sample of high quality CO and \carbi\ spectra with which to 
average out local velocity segregations which could mimic a change in any of the constants. 

\section*{Acknowledgements}
We would like to thank the referee Paolo Molaro for his prompt and helpful comments as well as
Stephen Lo for facilitating the enjoyable seafood lunch
overlooking Coogee Bay, during which the ideas for this paper were developed.
Also, Nick Tothill who came for coffee afterwards and threw in his two cents worth.
This
research has made use of the NASA/IPAC Extragalactic Database (NED)
which is operated by the Jet Propulsion Laboratory, California
Institute of Technology, under contract with the National Aeronautics
and Space Administration. This research has also made use of NASA's
Astrophysics Data System Bibliographic Service. 


\begin{thebibliography}{84}
\expandafter\ifx\csname natexlab\endcsname\relax\def\natexlab#1{#1}\fi

\bibitem[{{Ao} {et~al.}(2008){Ao}, {Wei{\ss}}, {Downes}, {Walter}, {Henkel}, \&
  {Menten}}]{awd+08}
{Ao}, Y., {Wei{\ss}}, A., {Downes}, D., {et~al.} 2008, A\&A, 491, 747

\bibitem[{{Barvainis} {et~al.}(1997){Barvainis}, {Maloney}, {Antonucci}, \&
  {Alloin}}]{bmaa97}
{Barvainis}, R., {Maloney}, P., {Antonucci}, R., \& {Alloin}, D. 1997, ApJ,
  484, 695

\bibitem[{Berengut {et~al.}(2011)Berengut, Flambaum, King, Curran, \&
  Webb}]{bfk+10}
Berengut, J.~C., Flambaum, V.~V., King, J.~A., Curran, S.~J., \& Webb, J.~K.
  2011, Phys. Rev. D, 83, 123506

\bibitem[{{Bertoldi} {et~al.}(2003){Bertoldi}, {Cox}, {Neri}, {Carilli},
  {Walter}, {Omont}, {Beelen}, {Henkel}, {Fan}, {Strauss}, \&
  {Menten}}]{bcn+03}
{Bertoldi}, F., {Cox}, P., {Neri}, R., {et~al.} 2003, A\&A, 409, L47

\bibitem[{{Blain}(1996)}]{bla96}
{Blain}, A.~W. 1996, MNRAS, 283, 1340

\bibitem[{{Blain} {et~al.}(1999){Blain}, {M\"{o}ller}, \& {Maller}}]{bmm99}
{Blain}, A.~W., {M\"{o}ller}, O., \& {Maller}, A.~H. 1999, MNRAS, 303, 423

\bibitem[{{Bothwell} {et~al.}(2010){Bothwell}, {Chapman}, {Tacconi}, {Smail},
  {Ivison}, {Casey}, {Bertoldi}, {Beswick}, {Biggs}, {Blain}, {Cox}, {Genzel},
  {Greve}, {Kennicutt}, {Muxlow}, {Neri}, \& {Omont}}]{bct+10}
{Bothwell}, M.~S., {Chapman}, S.~C., {Tacconi}, L., {et~al.} 2010, MNRAS, 405,
  219

\bibitem[{{Casey} {et~al.}(2011){Casey}, {Chapman}, {Neri}, {Bertoldi},
  {Smail}, {Greve}, {Beswick}, {Blain}, {Coppin}, {Cox}, {Genzel}, {Ivison},
  {Muxlow}, {Omont}, \& {Swinbank}}]{ccn+09}
{Casey}, C.~M., {Chapman}, S.~C., {Neri}, R., {et~al.} 2011, MNRAS, in press
  (arXiv:0910.5756)

\bibitem[{{Cooksy} {et~al.}(1986){Cooksy}, {Blake}, \& {Saykally}}]{cbs86}
{Cooksy}, A.~L., {Blake}, G.~A., \& {Saykally}, R.~J. 1986, ApJ, 305, L89

\bibitem[{{Coppin} {et~al.}(2007){Coppin}, {Swinbank}, {Neri}, {Cox}, {Smail},
  {Ellis}, {Geach}, {Siana}, {Teplitz}, {Dye}, {Kneib}, {Edge}, \&
  {Richard}}]{csn+07}
{Coppin}, K.~E.~K., {Swinbank}, A.~M., {Neri}, R., {et~al.} 2007, ApJ, 665, 936

\bibitem[{Cowie \& Songaila(1995)}]{cs95}
Cowie, L.~L. \& Songaila, A. 1995, ApJ, 453, 596

\bibitem[{{Cox} {et~al.}(2002){Cox}, {Omont}, {Djorgovski}, {Bertoldi}, {Pety},
  {Carilli}, {Isaak}, {Beelen}, {McMahon}, \& {Castro}}]{cod+02}
{Cox}, P., {Omont}, A., {Djorgovski}, S.~G., {et~al.} 2002, A\&A, 387, 406

\bibitem[{Curran(2000)}]{cur99p}
Curran, S.~J. 2000, A\&AS, 144, 271

\bibitem[{Curran(2009)}]{cur09}
Curran, S.~J. 2009, A\&A, 497, 351

\bibitem[{{Daddi} {et~al.}(2010){Daddi}, {Bournaud}, {Walter}, {Dannerbauer},
  {Carilli}, {Dickinson}, {Elbaz}, {Morrison}, {Riechers}, {Onodera}, {Salmi},
  {Krips}, \& {Stern}}]{dbw+10}
{Daddi}, E., {Bournaud}, F., {Walter}, F., {et~al.} 2010, ApJ, 713, 686

\bibitem[{{Daddi} {et~al.}(2009){Daddi}, {Dannerbauer}, {Stern}, {Dickinson},
  {Morrison}, {Elbaz}, {Giavalisco}, {Mancini}, {Pope}, \& {Spinrad}}]{dds+09}
{Daddi}, E., {Dannerbauer}, H., {Stern}, D., {et~al.} 2009, ApJ, 694, 1517

\bibitem[{{Danielson} {et~al.}(2011){Danielson}, {Swinbank}, {Smail}, {Cox},
  {Edge}, {Weiss}, {Harris}, {Baker}, {De Breuck}, {Geach}, {Ivison}, {Krips},
  {Lundgren}, {Longmore}, {Neri}, \& {Flaquer}}]{dss+11}
{Danielson}, A.~L.~R., {Swinbank}, A.~M., {Smail}, I., {et~al.} 2011, MNRAS,
  410, 1687

\bibitem[{{De Breuck} {et~al.}(2005){De Breuck}, {Downes}, {Neri}, {van
  Breugel}, {Reuland}, {Omont}, \& {Ivison}}]{ddn+05}
{De Breuck}, C., {Downes}, D., {Neri}, R., {et~al.} 2005, A\&A, 430, L1

\bibitem[{{De Breuck} {et~al.}(2003){De Breuck}, {Neri}, {Morganti}, {Omont},
  {Rocca-Volmerange}, {Stern}, {Reuland}, {van Breugel}, {R{\"o}ttgering},
  {Stanford}, {Spinrad}, {Vigotti}, \& {Wright}}]{dnm+03}
{De Breuck}, C., {Neri}, R., {Morganti}, R., {et~al.} 2003, A\&A, 401, 911

\bibitem[{{Downes} {et~al.}(1999){Downes}, {Neri}, {Wiklind}, {Wilner}, \&
  {Shaver}}]{dnw+99}
{Downes}, D., {Neri}, R., {Wiklind}, T., {Wilner}, D.~J., \& {Shaver}, P.~A.
  1999, ApJ, 513, L1

\bibitem[{{Downes} {et~al.}(1995){Downes}, {Solomon}, \& {Radford}}]{dsr95}
{Downes}, D., {Solomon}, P.~M., \& {Radford}, S.~J.~E. 1995, ApJ, 453, L65

\bibitem[{{Emonts} {et~al.}(2011){Emonts}, {Norris}, {Feain}, {Miley},
  {Sadler}, {Villar-Martin}, {Mao}, {Oosterloo}, {Ekers}, {Stevens},
  {Wieringa}, {Coppin}, \& {Tadhunter}}]{enf+11}
{Emonts}, B.~H.~C., {Norris}, R.~P., {Feain}, I., {et~al.} 2011, MNRAS, 415,
  655

\bibitem[{{Engargiola} {et~al.}(2003){Engargiola}, {Plambeck}, {Rosolowsky}, \&
  {Blitz}}]{eprb03}
{Engargiola}, G., {Plambeck}, R.~L., {Rosolowsky}, E., \& {Blitz}, L. 2003,
  ApJS, 149, 343

\bibitem[{{Frayer} {et~al.}(2011){Frayer}, {Harris}, {Baker}, {Ivison},
  {Smail}, {Negrello}, {Maddalena}, {Aretxaga}, {Baes}, {Birkinshaw},
  {Bonfield}, {Burgarella}, {Buttiglione}, {Cava}, {Clements}, {Cooray},
  {Dannerbauer}, {Dariush}, {De Zotti}, {Dunlop}, {Dunne}, {Dye}, {Eales},
  {Fritz}, {Gonzalez-Nuevo}, {Herranz}, {Hopwood}, {Hughes}, {Ibar}, {Jarvis},
  {Lagache}, {Leeuw}, {Lopez-Caniego}, {Maddox}, {Micha{\l}owski}, {Omont},
  {Pohlen}, {Rigby}, {Rodighiero}, {Scott}, {Serjeant}, {Smith}, {Swinbank},
  {Temi}, {Thompson}, {Valtchanov}, {van der Werf}, \& {Verma}}]{fhb+11}
{Frayer}, D.~T., {Harris}, A.~I., {Baker}, A.~J., {et~al.} 2011, ApJ, 726, L22

\bibitem[{{Frayer} {et~al.}(1999){Frayer}, {Ivison}, {Scoville}, {Evans},
  {Yun}, {Smail}, {Barger}, {Blain}, \& {Kneib}}]{fis+99}
{Frayer}, D.~T., {Ivison}, R.~J., {Scoville}, N.~Z., {et~al.} 1999, ApJ, 514,
  L13

\bibitem[{{Frayer} {et~al.}(1998){Frayer}, {Ivison}, {Scoville}, {Yun},
  {Evans}, {Smail}, {Blain}, \& {Kneib}}]{fis+98}
{Frayer}, D.~T., {Ivison}, R.~J., {Scoville}, N.~Z., {et~al.} 1998, ApJ, 506,
  L7

\bibitem[{{Garrett} {et~al.}(2005){Garrett}, {Knudsen}, \& {van der
  Werf}}]{gkv05}
{Garrett}, M.~A., {Knudsen}, K.~K., \& {van der Werf}, P.~P. 2005, A\&A, 431,
  L21

\bibitem[{{Genzel} {et~al.}(2003){Genzel}, {Baker}, {Tacconi}, {Lutz}, {Cox},
  {Guilloteau}, \& {Omont}}]{gbt+03}
{Genzel}, R., {Baker}, A.~J., {Tacconi}, L.~J., {et~al.} 2003, ApJ, 584, 633

\bibitem[{Gerin \& Phillips(2000)}]{gp00}
Gerin, M. \& Phillips, T.~G. 2000, ApJ, 537, 644

\bibitem[{{Greve} {et~al.}(2003){Greve}, {Ivison}, \& {Papadopoulos}}]{gip03}
{Greve}, T.~R., {Ivison}, R.~J., \& {Papadopoulos}, P.~P. 2003, ApJ, 599, 839

\bibitem[{{Hainline} {et~al.}(2004){Hainline}, {Scoville}, {Yun}, {Hawkins},
  {Frayer}, \& {Isaak}}]{hsy+04}
{Hainline}, L.~J., {Scoville}, N.~Z., {Yun}, M.~S., {et~al.} 2004, ApJ, 609, 61

\bibitem[{Henkel {et~al.}(2009)Henkel, Menten, Murphy, Jethava, Flambaum,
  Braatz, Muller, Ott, \& Mao}]{hmm+09}
Henkel, C., Menten, K.~M., Murphy, M.~T., {et~al.} 2009, A\&A, 500, 725

\bibitem[{{Ikeda} {et~al.}(2002){Ikeda}, {Oka}, {Tatematsu}, {Sekimoto}, \&
  {Yamamoto}}]{iot+02}
{Ikeda}, M., {Oka}, T., {Tatematsu}, K., {Sekimoto}, Y., \& {Yamamoto}, S.
  2002, ApJS, 139, 467

\bibitem[{{Iono} {et~al.}(2006{\natexlab{a}}){Iono}, {Tamura}, {Nakanishi},
  {Kawabe}, {Kohno}, {Okuda}, {Yamada}, {Hatsukade}, \& {Sameshima}}]{itn+06}
{Iono}, D., {Tamura}, Y., {Nakanishi}, K., {et~al.} 2006{\natexlab{a}}, PASJ,
  58, 957

\bibitem[{{Iono} {et~al.}(2006{\natexlab{b}}){Iono}, {Yun}, {Elvis}, {Peck},
  {Ho}, {Wilner}, {Hunter}, {Matsushita}, \& {Muller}}]{iye+06}
{Iono}, D., {Yun}, M.~S., {Elvis}, M., {et~al.} 2006{\natexlab{b}}, ApJ, 645,
  L97

\bibitem[{Israel \& Baas(2002)}]{ib02}
Israel, F.~P. \& Baas, F. 2002, A\&A, 383, 82

\bibitem[{{Ivison} {et~al.}(2010){Ivison}, {Swinbank}, {Swinyard}, {Smail},
  {Pearson}, {Rigopoulou}, {Polehampton}, {Baluteau}, {Barlow}, {Blain},
  {Bock}, {Clements}, {Coppin}, {Cooray}, {Danielson}, {Dwek}, {Edge},
  {Franceschini}, {Fulton}, {Glenn}, {Griffin}, {Isaak}, {Leeks}, {Lim},
  {Naylor}, {Oliver}, {Page}, {P{\'e}rez Fournon}, {Rowan-Robinson}, {Savini},
  {Scott}, {Spencer}, {Valtchanov}, {Vigroux}, \& {Wright}}]{iss+10}
{Ivison}, R.~J., {Swinbank}, A.~M., {Swinyard}, B., {et~al.} 2010, A\&A, 518,
  L35

\bibitem[{{Kanekar} {et~al.}(2010){Kanekar}, {Prochaska}, {Ellison}, \&
  {Chengalur}}]{kpec10}
{Kanekar}, N., {Prochaska}, J.~X., {Ellison}, S.~L., \& {Chengalur}, J.~N.
  2010, ApJ, 712, L148

\bibitem[{King {et~al.}(2011)King, Murphy, Ubachs, \& Webb}]{kmuw11}
King, J.~A., Murphy, M.~T., Ubachs, W., \& Webb, J.~K. 2011, MNRAS, accepted
  (arXiv:1106.5786)

\bibitem[{{Klamer} {et~al.}(2005){Klamer}, {Ekers}, {Sadler}, {Weiss},
  {Hunstead}, \& {De Breuck}}]{kes+05}
{Klamer}, I.~J., {Ekers}, R.~D., {Sadler}, E.~M., {et~al.} 2005, ApJ, 621, L1

\bibitem[{{Klein} {et~al.}(1998){Klein}, {Lewen}, {Schieder}, {Stutzki}, \&
  {Winnewisser}}]{kls+98}
{Klein}, H., {Lewen}, F., {Schieder}, R., {Stutzki}, J., \& {Winnewisser}, G.
  1998, ApJ, 494, L125

\bibitem[{{Kneib} {et~al.}(2005){Kneib}, {Neri}, {Smail}, {Blain}, {Sheth},
  {van der Werf}, \& {Knudsen}}]{kns+05}
{Kneib}, J., {Neri}, R., {Smail}, I., {et~al.} 2005, A\&A, 434, 819

\bibitem[{{Lestrade} {et~al.}(2010){Lestrade}, {Combes}, {Salom{\'e}}, {Omont},
  {Bertoldi}, {Andr{\'e}}, \& {Schneider}}]{lcs+10}
{Lestrade}, J.-F., {Combes}, F., {Salom{\'e}}, P., {et~al.} 2010, A\&A, 522, L4

\bibitem[{{Lestrade} {et~al.}(2009){Lestrade}, {Wyatt}, {Bertoldi}, {Menten},
  \& {Labaigt}}]{lwb+09}
{Lestrade}, J.-F., {Wyatt}, M.~C., {Bertoldi}, F., {Menten}, K.~M., \&
  {Labaigt}, G. 2009, A\&A, 506, 1455

\bibitem[{{Levshakov} {et~al.}(2006){Levshakov}, {Centuri{\'o}n}, {Molaro},
  D'Odorico, Reimers, Quast, \& Pollmannand}]{lcm+06}
{Levshakov}, S.~A., {Centuri{\'o}n}, M., {Molaro}, P., {et~al.} 2006, A\&A,
  449, 879

\bibitem[{{Levshakov} {et~al.}(2008){Levshakov}, {Reimers}, {Kozlov}, {Porsev},
  \& {Molaro}}]{lrk+08}
{Levshakov}, S.~A., {Reimers}, D., {Kozlov}, M.~G., {Porsev}, S.~G., \&
  {Molaro}, P. 2008, A\&A, 479, 719

\bibitem[{{Lupu} {et~al.}(2011){Lupu}, {Scott}, {Aguirre}, {Aretxaga}, {Auld},
  {Barton}, {Beelen}, {Bertoldi}, {Bock}, {Bonfield}, {Bradford},
  {Buttiglione}, {Cava}, {Clements}, {Cooke}, {Cooray}, {Dannerbauer},
  {Dariush}, {De Zotti}, {Dunne}, {Dye}, {Eales}, {Frayer}, {Fritz}, {Glenn},
  {Hughes}, {Ibar}, {Ivison}, {Jarvis}, {Kamenetzky}, {Kim}, {Lagache},
  {Leeuw}, {Maddox}, {Maloney}, {Matsuhara}, {Murphy}, {Naylor}, {Negrello},
  {Nguien}, {Omont}, {Pascale}, {Pohlen}, {Rigby}, {Rodighiero}, {Serjeant},
  {Smith}, {Temi}, {Thompson}, {Valtchanov}, {Verma}, {Vieira}, \&
  {Zmuidzinas}}]{lsa+11}
{Lupu}, R.~E., {Scott}, K.~S., {Aguirre}, J.~E., {et~al.} 2011, ApJ, submitted
  (arXiv:1009.5983)

\bibitem[{{Maiolino} {et~al.}(2005){Maiolino}, {Cox}, {Caselli}, {Beelen},
  {Bertoldi}, {Carilli}, {Kaufman}, {Menten}, {Nagao}, {Omont}, {Wei{\ss}},
  {Walmsley}, \& {Walter}}]{mcc+05}
{Maiolino}, R., {Cox}, P., {Caselli}, P., {et~al.} 2005, A\&A, 440, L51

\bibitem[{{Maiolino} {et~al.}(2007){Maiolino}, {Neri}, {Beelen}, {Bertoldi},
  {Carilli}, {Caselli}, {Cox}, {Menten}, {Nagao}, {Omont}, {Walmsley},
  {Walter}, \& {Wei{\ss}}}]{mnb+07}
{Maiolino}, R., {Neri}, R., {Beelen}, A., {et~al.} 2007, A\&A, 472, L33

\bibitem[{{Muller} {et~al.}(2011){Muller}, {Beelen}, {Gu{\'e}lin}, {Aalto},
  {Black}, {Combes}, {Curran}, {Theule}, \& {Longmore}}]{mbg+11}
{Muller}, S., {Beelen}, A., {Gu{\'e}lin}, M., {et~al.} 2011, A\&A, submitted
  (arXiv:1104.3361)

\bibitem[{Murphy {et~al.}(2003)Murphy, Webb, \& Flambaum}]{mwf03}
Murphy, M.~T., Webb, J.~K., \& Flambaum, V.~V. 2003, MNRAS, 345, 609

\bibitem[{Murphy {et~al.}(2001)Murphy, Webb, Flambaum, Drinkwater, Combes, \&
  Wiklind}]{mwf+00}
Murphy, M.~T., Webb, J.~K., Flambaum, V.~V., {et~al.} 2001, MNRAS, 327, 1244

\bibitem[{{Ofek} {et~al.}(2003){Ofek}, {Rix}, \& {Maoz}}]{orm03}
{Ofek}, E.~O., {Rix}, H., \& {Maoz}, D. 2003, MNRAS, 343, 639

\bibitem[{{Ohta} {et~al.}(1996){Ohta}, {Yamada}, {Nakanishi}, {Kohno},
  {Akiyama}, \& {Kawabe}}]{oyn+96}
{Ohta}, K., {Yamada}, T., {Nakanishi}, K., {et~al.} 1996, Nat, 382, 426

\bibitem[{{Ojha} {et~al.}(2001){Ojha}, {Stark}, {Hsieh}, {Lane}, {Chamberlin},
  {Bania}, {Bolatto}, {Jackson}, \& {Wright}}]{osh+01}
{Ojha}, R., {Stark}, A.~A., {Hsieh}, H.~H., {et~al.} 2001, ApJ, 548, 253

\bibitem[{{Omont} {et~al.}(1996){Omont}, {Petitjean}, {Guilloteau}, {McMahon},
  {Solomon}, \& {P{\'e}contal}}]{opg+96}
{Omont}, A., {Petitjean}, P., {Guilloteau}, S., {et~al.} 1996, Nat, 382, 428

\bibitem[{{Papadopoulos} \& {Greve}(2004)}]{pg04}
{Papadopoulos}, P.~P. \& {Greve}, T.~R. 2004, ApJ, 615, L29

\bibitem[{{Papadopoulos} {et~al.}(2000){Papadopoulos}, {R{\"o}ttgering}, {van
  der Werf}, {Guilloteau}, {Omont}, {van Breugel}, \& {Tilanus}}]{prv+00}
{Papadopoulos}, P.~P., {R{\"o}ttgering}, H.~J.~A., {van der Werf}, P.~P.,
  {et~al.} 2000, ApJ, 528, 626

\bibitem[{{Papadopoulos} {et~al.}(2004){Papadopoulos}, {Thi}, \&
  {Viti}}]{ptv04}
{Papadopoulos}, P.~P., {Thi}, W.-F., \& {Viti}, S. 2004, MNRAS, 351, 147

\bibitem[{{Pety} {et~al.}(2004){Pety}, {Beelen}, {Cox}, {Downes}, {Omont},
  {Bertoldi}, \& {Carilli}}]{pbc+04}
{Pety}, J., {Beelen}, A., {Cox}, P., {et~al.} 2004, A\&A, 428, L21

\bibitem[{Press {et~al.}(1989)Press, Flannery, Teukolsky, \&
  Vetterling}]{pftv89}
Press, W.~H., Flannery, B.~P., Teukolsky, S.~A., \& Vetterling, W.~T. 1989,
  Numerical Recipes: {T}he Art of Scientific Computing (Cambridge: Cambridge
  University Press)

\bibitem[{{Riechers}(2011)}]{rie11a}
{Riechers}, D.~A. 2011, ApJ, 730, 108

\bibitem[{{Riechers} {et~al.}(2010){Riechers}, {Capak}, {Carilli}, {Cox},
  {Neri}, {Scoville}, {Schinnerer}, {Bertoldi}, \& {Yan}}]{rcc+10}
{Riechers}, D.~A., {Capak}, P.~L., {Carilli}, C.~L., {et~al.} 2010, ApJ, 720,
  L131

\bibitem[{{Riechers} {et~al.}(2009){Riechers}, {Walter}, {Carilli}, \&
  {Lewis}}]{rwcl09}
{Riechers}, D.~A., {Walter}, F., {Carilli}, C.~L., \& {Lewis}, G.~F. 2009, ApJ,
  690, 463

\bibitem[{{Scoville} {et~al.}(1997){Scoville}, {Yun}, {Windhorst}, {Keel}, \&
  {Armus}}]{syw+97}
{Scoville}, N.~Z., {Yun}, M.~S., {Windhorst}, R.~A., {Keel}, W.~C., \& {Armus},
  L. 1997, ApJ, 485, L21

\bibitem[{{Sheth} {et~al.}(2008){Sheth}, {Vogel}, {Wilson}, \& {Dame}}]{svwd08}
{Sheth}, K., {Vogel}, S.~N., {Wilson}, C.~D., \& {Dame}, T.~M. 2008, ApJ, 675,
  330

\bibitem[{{Solomon} \& {Vanden Bout}(2005)}]{sv05}
{Solomon}, P.~M. \& {Vanden Bout}, P.~A. 2005, Ann. Rev. Astr. Ap., 43, 677

\bibitem[{{Spergel} {et~al.}(2003){Spergel}, {Verde}, {Peiris}, {Komatsu},
  {Nolta}, {Bennett}, {Halpern}, {Hinshaw}, {Jarosik}, {Kogut}, {Limon},
  {Meyer}, {Page}, {Tucker}, {Weiland}, {Wollack}, \& {Wright}}]{svp+03}
{Spergel}, D.~N., {Verde}, L., {Peiris}, H.~V., {et~al.} 2003, ApJS, 148, 175

\bibitem[{{Stutzki} {et~al.}(1997){Stutzki}, {Graf}, {Haas}, {Honingh},
  {Hottgenroth}, {Jacobs}, {Schieder}, {Simon}, {Staguhn}, {Winnewisser},
  {Martin}, {Peters}, \& {McMullin}}]{sgh+97}
{Stutzki}, J., {Graf}, U.~U., {Haas}, S., {et~al.} 1997, ApJ, 477, L33

\bibitem[{{Tacconi} {et~al.}(2010){Tacconi}, {Genzel}, {Neri}, {Cox}, {Cooper},
  {Shapiro}, {Bolatto}, {Bouch{\'e}}, {Bournaud}, {Burkert}, {Combes},
  {Comerford}, {Davis}, {Schreiber}, {Garcia-Burillo}, {Gracia-Carpio}, {Lutz},
  {Naab}, {Omont}, {Shapley}, {Sternberg}, \& {Weiner}}]{tgn+10}
{Tacconi}, L.~J., {Genzel}, R., {Neri}, R., {et~al.} 2010, Nat, 463, 781

\bibitem[{{Tacconi} {et~al.}(2006){Tacconi}, {Neri}, {Chapman}, {Genzel},
  {Smail}, {Ivison}, {Bertoldi}, {Blain}, {Cox}, {Greve}, \& {Omont}}]{tnc+06}
{Tacconi}, L.~J., {Neri}, R., {Chapman}, S.~C., {et~al.} 2006, ApJ, 640, 228

\bibitem[{{Uzan}(2003)}]{uza03}
{Uzan}, J. 2003, Reviews of Modern Physics, 75, 403

\bibitem[{{Wagg} {et~al.}(2006){Wagg}, {Wilner}, {Neri}, {Downes}, \&
  {Wiklind}}]{wwn+06}
{Wagg}, J., {Wilner}, D.~J., {Neri}, R., {Downes}, D., \& {Wiklind}, T. 2006,
  ApJ, 46

\bibitem[{{Walter} {et~al.}(2003){Walter}, {Bertoldi}, {Carilli}, {Cox}, {Lo},
  {Neri}, {Fan}, {Omont}, {Strauss}, \& {Menten}}]{wbc+03}
{Walter}, F., {Bertoldi}, F., {Carilli}, C., {et~al.} 2003, Nat, 424, 406

\bibitem[{{Walter} {et~al.}(2011){Walter}, {Wei{\ss}}, {Downes}, {Decarli}, \&
  {Henkel}}]{wwd+11}
{Walter}, F., {Wei{\ss}}, A., {Downes}, D., {Decarli}, R., \& {Henkel}, C.
  2011, ApJ, 730, 18

\bibitem[{Webb {et~al.}(2011)Webb, King, Murphy, Flambaum, Carswell, \&
  Bainbridge}]{wkm+10}
Webb, J.~K., King, J.~A., Murphy, M.~T., {et~al.} 2011, PhRvL, submitted
  (arXiv:1008.3907)

\bibitem[{{Wei{\ss}} {et~al.}(2005{\natexlab{a}}){Wei{\ss}}, {Downes},
  {Henkel}, \& {Walter}}]{wdhw05}
{Wei{\ss}}, A., {Downes}, D., {Henkel}, C., \& {Walter}, F. 2005{\natexlab{a}},
  A\&A, 429, L25

\bibitem[{{Wei{\ss}} {et~al.}(2005{\natexlab{b}}){Wei{\ss}}, {Downes},
  {Walter}, \& {Henkel}}]{wdwh05}
{Wei{\ss}}, A., {Downes}, D., {Walter}, F., \& {Henkel}, C. 2005{\natexlab{b}},
  A\&A, 440, L45

\bibitem[{{Wei{\ss}} {et~al.}(2003){Wei{\ss}}, {Henkel}, {Downes}, \&
  {Walter}}]{whdw03}
{Wei{\ss}}, A., {Henkel}, C., {Downes}, D., \& {Walter}, F. 2003, A\&A, 409

\bibitem[{Wei{\ss} {et~al.}(2009)Wei{\ss}, Ivison, Downes, Walter, Cirasuolo,
  \& Menten}]{wid+09}
Wei{\ss}, A., Ivison, R.~J., Downes, D., {et~al.} 2009, ApJ, 705, L45

\bibitem[{{White} {et~al.}(1994){White}, {Ellison}, {Claude}, {Dent}, \&
  {Matheson}}]{wec+94}
{White}, G.~J., {Ellison}, B., {Claude}, S., {Dent}, W. R.~F., \& {Matheson},
  D.~N. 1994, A\&A, 284, L23

\bibitem[{{Willott} {et~al.}(2007){Willott}, {Mart{\'{\i}}nez-Sansigre}, \&
  {Rawlings}}]{wmr07}
{Willott}, C.~J., {Mart{\'{\i}}nez-Sansigre}, A., \& {Rawlings}, S. 2007, AJ,
  133, 564

\bibitem[{{Yamamoto} \& {Saito}(1991)}]{ys91a}
{Yamamoto}, S. \& {Saito}, S. 1991, ApJ, 370, L103

\bibitem[{{Yan} {et~al.}(2010){Yan}, {Tacconi}, {Fiolet}, {Sajina}, {Omont},
  {Lutz}, {Zamojski}, {Neri}, {Cox}, \& {Dasyra}}]{ytf+10}
{Yan}, L., {Tacconi}, L.~J., {Fiolet}, N., {et~al.} 2010, ApJ, 714, 100

\end{thebibliography}

\end{document}